\documentclass{aa}

\usepackage{graphicx}
\usepackage{txfonts}
\usepackage{hyperref}
\begin{document}

    \title{First models of \textit{s} process in AGB stars of solar metallicity for the stellar evolutionary code ATON with a novel stable explicit numerical solver}
    \titlerunning{First \textit{s} process AGB ATON models with a novel stable explicit numerical solver}
   
    \author{
        A. Yag\"ue L\'opez\inst{1, 9}
        \and
        D. A. Garc\'ia-Hern\'andez\inst{2, 3}
        \and
        P. Ventura\inst{4}
        \and
        C. L. Doherty\inst{1, 5}
        \and
        J. W. den Hartogh\inst{1, 6, 9}
        \and
        S. W. Jones\inst{7, 8, 9}
        \and
        M. Lugaro\inst{1, 5, 6}
        }

   \institute{Konkoly Observatory, Research Centre for Astronomy and Earth Sciences, E\"otv\"os Lor\'and Research Network (ELKH), Konkoly Thege Mikl\'{o}s \'{u}t 15-17, H-1121 Budapest, Hungary\\
        \email{andres.yague@csfk.org}
        \and
        Instituto de Astrof\'isica de Canarias, C/Via L\'actea s/n, E-38205 La Laguna, Spain
        \and
        Departamento de Astrof\'isica, Universidad de La Laguna (ULL), E-38206 La Laguna, Spain
        \and
        INAF-Osservatorio Astronomico di Roma, via Frascati 33, 00040 Monteporzio, Italy
        \and
        School of Physics and Astronomy, Monash University, VIC 3800, Australia
        \and
        ELTE E\"{o}tv\"{o}s Lor\'and University, Institute of Physics, Budapest 1117, P\'azm\'any P\'eter s\'et\'any 1/A, Hungary
        \and
        X Computational Physics (XCP) Division, Los Alamos National Laboratory, Los Alamos, NM 87545, USA
        \and
        Computer, Computational and Statistical Sciences (CCS) Division, Los Alamos National Laboratory, Los Alamos, NM 87545, USA
        \and
        Nugrid collaboration
        }

   \date{}

 
  \abstract
   {}
    {We describe the first \textit{s}-process post-processing models for asymptotic giant branch (AGB) stars of masses 3, 4 and 5 M$_\odot$ at solar metallicity (Z = 0.018) computed using the input from the stellar evolutionary code \textsc{aton}.}
    {The models are computed with the new code \textsc{snuppat} (\textit{S}-process NUcleosynthesis Post-Processing code for \textsc{aton}), including an advective scheme for the convective overshoot that leads to the formation of the main neutron source, $^{13}$C. Each model is post-processed with 3 different values of the free overshoot parameter. Included in the code \textsc{snuppat} is the novel Patankar-Euler-Deflhard explicit numerical solver, that we use to solve the nuclear network system of differential equations.}
    {The results are compared to those from other \textit{s}-process nucleosynthesis codes (Monash, \textsc{fruity}, and NuGrid), as well as observations of \textit{s}-process enhancement in AGB stars, planetary nebulae, and barium stars. This comparison shows that the relatively high abundance of $^{12}$C in the He-rich intershell in \textsc{aton} results in a \textit{s}-process abundance pattern that favours the second over the first \textit{s}-process peak for all the masses explored. Also, our choice of an advective as opposed to diffusive numerical scheme for the convective overshoot results in significant \textit{s}-process nucleosynthesis also for the 5 M$_\odot$ models, which may be in contradiction with observations.}
   {}


   \maketitle
%

\section{Introduction}

    Low- to intermediate-mass (initial mass in the range $0.8$ M$_\odot$ $\le$ M $\le 8$ M$_\odot$) thermally pulsing asymptotic giant branch (AGB, \citealp{Iben1983}) stars are predicted to contribute $\sim50\%$ of the cosmic abundances of the elements heavier than iron in the Galaxy (e.g., \citealp{Arlandini1999}). These late stage stars are structured as an inert CO core surrounded by a He-rich region (known as the He intershell) contained inside a H-rich convective envelope. Energy generation is alternately activated in two sites: the bottom of the He intershell through He burning, and the H-burning shell through the CNO cycle. Depending on the active burning process, the star is found to be either in a thermal pulse (TP, He-burning phase) or an interpulse period (H-burning phase). There are two important changes in the stellar structure when the star goes through a TP (e.g., \citealp{Herwig2000}). The first is the formation in the He intershell of a convective zone known as the pulse-driven convective zone (PDCZ). The second, which takes place after the PDCZ disappears, is the penetration of the H-rich convective envelope into the He intershell, a process known as the third dredge-up (TDU), which allows the products of He burning to reach the stellar surface.
    
    Heavy element nucleosynthesis via the \textit{slow} neutron-capture process (\textit{s} process, e.g., \citealp{Clayton1961}) occurs in the He intershell of AGB stars (\citealp{Schwarzschild1967, Sanders1967, Iben1975A}). The \textit{s} process consists of a chain of neutron captures and $\beta$-decays that increase the heavy nuclei abundances along the valley of stability. The \textit{s} process takes place in the He intershell when enough free neutrons are present due to the activation of two neutron source reactions: $^{13}$C($\alpha$,n)$^{16}$O and $^{22}$Ne($\alpha$,n)$^{25}$Mg (see Sect.~\ref{sec:Sprocess} for a more detailed description). During the TDU events the products of the \textit{s} process are mixed up to the stellar surface, where they can be observed and they affect the chemical evolution of the Galaxy because they are ejected by the strong stellar winds that develop during the AGB (for a comprehensive review see \citealp{Karakas2014}).
    
    To understand both galactic and AGB chemical evolution, it is essential to study the \textit{s} process with the help of numerical simulations that use information on AGB stellar structure evolution to calculate the \textit{s}-process nucleosynthesis. Owing to the computationally expensive solution of the nucleosynthesis equations, a common approach (e.g., \citealp{Karakas2016, Pignatari2016}) is to divide the calculation in two steps. In the first step is to calculate the relatively computationally cheap stellar structure evolution with a nuclear network containing only the most relevant reactions from an energetic point of view. The second step is to run a post-processing code using the stellar structure as an input and a more complete nuclear network capable of following the \textit{s}-process nucleosynthesis. On top of temperatures and densities, information about the location and extent of the convective regions is needed to correctly simulate the PDCZ and the effect of TDU on the surface stellar abundances. A less common, but more realistic and computationally expensive alternative is to solve the nucleosynthesis and structural equations simultaneously (e.g., \citealp{Cristallo2015}).
    
    A traditional approach to solve a nuclear network is to use an implicit or semi-implicit numerical solver \citep{Longland2014}, that is, an integration method that uses information about future abundances to solve for those same future abundances. The advantages of such solvers is that they are numerically stable, keeping the future abundances bounded between a maximum and minimum value if the analytical solution is expected to be bounded as well \citep[see, e.g.][]{Bader1983}. The main disadvantage is that these solvers require the solution to an algebraic system of equations, which is computationally costly. In order to sidestep this disadvantage, we present the explicit Patankar-Euler-Deuflhard (PED) solver, that has the advantage of being numerically stable and does not require the solution to an algebraic system of equations.
    
    Different AGB stellar evolutionary codes adopt different prescriptions to describe the physical processes inside AGB stars. Consequently, the predicted stellar structure evolution differs from code to code, affecting the nucleosynthesis calculations. In particular, the stellar structure evolutionary code \textsc{aton} (\citealp{Ventura1998, Ventura2008}) stands apart in that it uses the Full Spectrum of Turbulence (FST) (\citealp{Canuto1996}), a model for convection that takes into account the full spectrum of the eddies in the evaluation of the convective flux. Most stellar evolutionary codes use instead the Mixing Length Theory (MLT) (\citealp{Boehm1958}) scheme for convection, which takes into account only one large eddy. The use of the FST scheme along with other physical choices, such as the stellar wind prescription or the equation of state in the pressure ionisation regime, produce notable differences in several nucleosynthesis mechanisms. Examples of such mechanisms are the Hot Bottom Burning strength (HBB, that is, proton burning at the bottom of the convective envelope, see \citealp{Mazzitelli1999}), the number of thermal pulses (TP), and the maximum He intershell temperature. To compare and analyse the \textit{s}-process nucleosynthesis predictions obtained from the stellar structure provided by \textsc{aton} to those calculated with other evolutionary codes can help us test the validity of the simulations of both the stellar evolutionary and nucleosynthesis processes. However, while there are \textit{s}-process nucleosynthesis models for the codes that use the MLT convection scheme, no  such calculations have been carried out for \textsc{aton}, limiting its predictive capabilities (e.g., \citealp{GarciaHernandez2013}). With these calculations as our objective we have developed the \textsc{snuppat} code (\textit{S}-process NUcleosynthesis Post-Processing code for \textsc{aton}) described in this paper.
    
    The paper is structured as follows: a brief introduction to the \textit{s}-process nucleosynthesis can be found in Sect.~\ref{sec:Sprocess}. The methodology of this work is split in Sect.~\ref{sec:ATON}, which presents the \textsc{aton} models used as input for our post-processing code \textsc{snuppat}, and Sect.~\ref{sec:Snuppat}, which describes the post-processing code itself, including the novel explicit numerical solver used in this work. The results are presented in Sect.~\ref{sec:Results}, which deals with the formation of the main neutron source $^{13}$C in the post-processing code and presents the stellar surface abundances for a range of simulation parameters. Finally, the results are discussed and compared to other codes and observations in Sect.~\ref{sec:Discussion}, and the final remarks and future work can be found in Sect.~\ref{sec:Conclusions}.

\section{\textit{s}-Process nucleosynthesis in AGB stars}
\label{sec:Sprocess}

    As mentioned above, two neutron sources in the He intershell provide the free neutrons needed for the \textit{s} process. The first is the $^{13}$C($\alpha$,n)$^{16}$O reaction (e.g., \citealp{Straniero1995, Gallino1998, Abia2001}). This reaction takes place in a $^{13}$C-rich region known as the \textit{$^{13}$C pocket} that forms when enough protons from the H-rich convective envelope are mixed into the $^{12}$C-rich He intershell during the deepest extent of the TDU through the $^{12}$C(p,$\gamma$)$^{13}$N($\beta^+$,$\nu$)$^{13}$C reaction. This $^{13}$C burns radiatively during the interpulse period that lasts thousands to tens of thousands of years, generating a neutron density $\sim 10^8$cm$^{-3}$. Its nucleosynthesis products are then engulfed into the PDCZ. Although the exact mechanisms of the proton mixing are unknown, several possibilities have been invoked, such as convective overshoot (e.g., \citealp{Herwig1997, Herwig2000, Cristallo2009}), rotation (e.g., \citealp{Herwig2003}), gravity waves (\citealp{Denissenkov2003}), and magnetic buoyancy (\citealp{Trippella2016}). This proton mixing must present a decreasing profile with a shallow enough slope (i.e., it must produce a \textit{partial mixing zone}, see \citealp{Goriely2000, Lugaro2003, Cristallo2009, Buntain2017}) to result in a significant production of \textit{s}-process elements. This is because where too many protons are mixed, $^{13}$C is destroyed producing the neutron poison $^{14}$N, through $^{13}$C(p,$\gamma$)$^{14}$N, which captures the neutrons generated by the remaining $^{13}$C and inhibits the \textit{s}-process nucleosynthesis. It is useful to define the quantity
    \begin{equation}
        X_{^{13}\text{C}_\text{eff}} = X_{^{13}\text{C}} - \frac{13}{14}X_{^{14}\text{N}}
        \label{eq:effPocket}
    \end{equation}
    which, where greater than zero, is known as the \textit{effective} $^{13}$C pocket (see \citealp{Cristallo2009}) and defines the region in which the \textit{s}-process nucleosynthesis takes place for the $^{13}$C neutron source. Additionally, it has been predicted (\citealp{Goriely2004}) that under a diffusive mixing approach, if the temperature at the bottom of the convective envelope where the $^{13}$C pocket forms is above $\sim 70$ MK during the deepest extent of the TDU (known as a ``hot TDU''), the $^{13}$C pocket is always contained inside a $^{14}$N pocket, inhibiting this neutron source altogether. These hot TDUs can be found to occur in initial masses above $\sim 4$ M$_\odot$ for solar metallicity and down to $\sim 3$ M$_\odot$ for lower metallicities.
    
    The addition of a chemical mixing mechanism through rotation can also inhibit the $^{13}$C neutron source due to potential overlap of the $^{13}$C and $^{14}$N pockets. The main difference between this kind of inhibition and the one predicted in \citet{Goriely2004}, besides the intrinsic physical mechanism, is that chemical mixing through rotation can inhibit the $^{13}$C neutron source even in low mass stars (\citealp{Herwig2003, Siess2004, Piersanti2013}), which contradicts the observational data of \textit{s}-process enhanced objects such as Ba stars (\citealp{Cseh2018}). Moreover, recent simulations compared to asteroseismologic observations of the core rotation of giant stars indicate that rotation might not affect the \textit{s}-process nucleosynthesis in a significant way (\citealp{Hartogh2019}).
    
    The second neutron source is the $^{22}$Ne($\alpha$,n)$^{25}$Mg reaction (e.g., \citealp{Cameron1960, Iben1975A, Iben1975B, Busso1999, GarciaHernandez2006}). This reaction occurs at the bottom of the PDCZ, which is rich in the $^{22}$Ne produced by two successive $\alpha$ captures on the $^{14}$N ingested from the H-burning ashes. This neutron source is limited by its high activation temperature of $\gtrsim 300$ MK, restricting it to the more massive ($> 3$ M$_\odot$) of the intermediate-mass AGB stars (e.g., \citealp{GarciaHernandez2006, GarciaHernandez2009}), and by the timescale of the PDCZ. Contrary to the $^{13}$C neutron source, the $^{22}$Ne neutron source acts in a much shorter timescale of decades, but provides a significantly higher neutron density of up to $10^{13}$ cm$^{-3}$ (\citealp{vanRaai2012, Fishlock2014}). Another key difference is that, because the $^{22}$Ne neutron source takes place at the bottom of the PDCZ, the enriched material is replenished by fresh material from the rest of the PDCZ, effectively further lowering the neutron exposure.
    
    These two neutron sources produce markedly different \textit{s}-process abundance patterns. In particular, at solar metallicity the $^{22}$Ne neutron source produces the elements of the first \textit{s}-process peak (elements with a neutron magic number of 50, such as Sr, Y, or Zr, also known as the \textit{light} \textit{s}-process elements) and increases certain elemental ratios such as [Rb/Sr]\footnote{The bracket notation [A/B] is defined as $\log\left(Y_\text{A}/Y_\text{B}\right) - \log\left(Y_\text{A}/Y_\text{B}\right)_\odot$, where $Y_\text{A}$ is the number fraction of element A.}. The $^{13}$C neutron source produces elements of the second (magic number of neutrons of 82, such as Ba, La, or Ce, also known as the \textit{heavy} \textit{s}-process elements) and third (Pb) \textit{s}-process peaks. There are two reasons for these unique signatures: 1) the different neutron densities, which activate different paths due to the possible activation of the \textit{s}-process branching points, and 2) the time span during which each neutron source reaction is active, which changes the neutron exposure and, therefore, the relative production of the \textit{s}-process peaks. We briefly review both mechanisms.
    
    1) Branching points are nuclei on the \textit{s}-process path for which the fastest timescale can switch from the $\beta$-decay to the neutron capture reaction. A well known example of a branching point is $^{86}$Rb (e.g., \citealp{vanRaai2012}), for which the probability of decaying before capturing a neutron increases from $2.58\%$ for a neutron density of $10^8$ cm$^{-3}$ to $99.96\%$ for a neutron density of $10^{13}$ cm$^{-3}$. This branching point, along with that at $^{85}$Kr, leads to the creation of $^{87}$Rb, a long-lived ($\sim 10^{10}$ years) Rb isotope with a magic number of 50 neutrons (e.g., \citealp{GarciaHernandez2006}). Because the neutron densities necessary to activate these branching points are restricted to the $^{22}$Ne neutron source, their activation signals the presence of this neutron source in AGB stars. Therefore, the abundance of Rb is an indicator of the neutron density attained in the stellar interior. A similar situation can be found for $^{135}$Cs (half-life of $\sim 10^6$ years), which can only be produced from the stable $^{133}$Cs by neutron captures on the $^{134}$Cs (half-life of $\sim 2$ years).
    
    2) The neutron exposure $\tau$, defined as
    \begin{equation}
        \tau = \int_0^t N_n u_T dt',
        \label{eq:neutronExposure}
    \end{equation}
    where $N_n$ is the neutron density and $u_T$ is the thermal velocity, is the main predictor of the overall elemental \textit{s}-process abundance distribution. Values below $\sim 0.4$ mbarn$^{-1}$ (calculated for a temperature of about $100$ MK, see Fig.~2 of \citealp{Herwig2003}), heavily favour the first \textit{s}-process peak over the second and third, while a higher neutron exposure increases the production of the second and third peaks instead. The reason is that due to the small neutron capture cross sections of isotopes with a magic number of neutrons, the \textit{s}-process peaks act as bottlenecks on the \textit{s}-process path, halting it until the abundance build-up of these isotopes brings the neutron capture probability on par with that of the nuclei before the peak. An increase of either the neutron density or the time that neutron-capture reactions are active increases the neutron exposure, enhancing the production of heavy elements with respect to lower neutron exposures.

\section{Methods and models}

    As stated in the Introduction, our approach involves using the \textsc{aton} stellar evolutionary code and a post-processing code that uses the \textsc{aton} results as an input to accurately follow the \textit{s}-process nucleosynthesis.
    
    \subsection{AGB modelling with FST}
    \label{sec:ATON}
    
    We used three models with initial masses of 3, 4, and 5 M$_\odot$ and a metallicity of $Z = 0.018$ (scaled solar from \citealp{Asplund2009}) calculated with \textsc{aton}. These models were run with 30 species from H to Si, those necessary to correctly follow the most important reactions for the stellar evolution, with the cross sections taken from the NACRE compilation (\citealp{Angulo1999}). For all the models we used the FST prescription for convection, with the fine tuning parameter $\beta$ set to $0.2$, consistent with previous works (e.g., \citealp{Ventura1998, Mazzitelli1999}). Regarding overshoot of the convective eddies into radiative stable regions, we assumed that convective velocities decay from the formal convective/radiative interface with an e-folding distance of $\zeta H_{\text{P}}$, where $\zeta = 0.002$; this choice is in agreement with the calibration based on the luminosity function of C-stars in the LMC given in \citet{Ventura2014}. For the mass loss prescription, we have used the Bl\"ocker mass loss with a Reimers parameter of $0.02$ (\citealp{Bloecker1995}). Some of the stellar structure features are presented in Tables \ref{tab:3MSunATON}, \ref{tab:4MSunATON} and \ref{tab:5MSunATON}, respectively, for the 3, 4 and 5 M$_\odot$ stars.
    
    These models have been chosen because each one represents a distinctly different regime from a nucleosynthesis perspective: The 3 M$_\odot$ model surface composition is mostly changed due to TDU episodes and the presence of the $^{13}$C neutron source. The 5 M$_\odot$ model surface composition is affected by a strong HBB and $^{22}$Ne neutron source activation. Lastly, the 4 M$_\odot$ model sits in between, with a weaker HBB and $^{22}$Ne neutron source activation.
    
    \begin{table}
        \centering
        \caption{Physical characteristics related to the TDU events taking place in a 3 M$_\odot$ \textsc{aton} simulation with a convective envelope overshoot parameter of $\omega = 0.002$ (described in Sec.~\ref{sec:Snuppat}). The maximum intershell temperature ($T_\text{max}$) and the temperature at the bottom of the convective envelope during the deepest extent of the TDU ($T_\text{TDU}$) are expressed in MK. The core mass ($M_\text{core}$), envelope mass ($M_\text{env}$), and dredged-up mass ($M_\text{dredge}$) are presented in M$_\odot$. The criterion for counting a TDU in this table is that $M_\text{dredge} > 10^{-5}$}.
        \label{tab:3MSunATON}
        \begin{tabular}{lcccccr}
            \hline
            TDU \# & $\lambda$ & $T_\text{max}$ & $T_\text{TDU}$ & $M_\text{core}$ & $M_\text{env}$ & $M_\text{dredge}$\\
            \hline
            1 & 0.21 & 279 & 62 & 0.612 & 2.367 & 1.35e-03 \\
            2 & 0.27 & 287 & 63 & 0.617 & 2.354 & 1.87e-03 \\
            3 & 0.33 & 289 & 64 & 0.621 & 2.340 & 2.41e-03 \\
            4 & 0.37 & 289 & 64 & 0.626 & 2.325 & 2.72e-03 \\
            5 & 0.41 & 295 & 65 & 0.630 & 2.308 & 3.06e-03 \\
            6 & 0.45 & 295 & 65 & 0.635 & 2.289 & 3.35e-03 \\
            7 & 0.47 & 295 & 66 & 0.639 & 2.268 & 3.55e-03 \\
            8 & 0.50 & 292 & 66 & 0.642 & 2.246 & 3.77e-03 \\
            9 & 0.49 & 299 & 66 & 0.646 & 2.221 & 3.68e-03 \\
            10 & 0.51 & 293 & 66 & 0.650 & 2.195 & 3.79e-03 \\
            11 & 0.51 & 308 & 67 & 0.653 & 2.167 & 3.75e-03 \\
            12 & 0.54 & 298 & 67 & 0.657 & 2.112 & 3.97e-03 \\
            \hline
        \end{tabular}
    \end{table}
    
    \begin{table}
        \centering
        \caption{Same as Table~\ref{tab:3MSunATON} for a 4 M$_\odot$ simulation.}
        \label{tab:4MSunATON}
        \begin{tabular}{lcccccr}
            \hline
            TDU \# & $\lambda$ & $T_\text{max}$ & $T_\text{TDU}$ & $M_\text{core}$ & $M_\text{env}$ & $M_\text{dredge}$\\
            \hline
            1 & 0.01 & 293 & 76 & 0.804 & 3.063 & 2.90e-05 \\
            2 & 0.08 & 291 & 77 & 0.806 & 3.030 & 2.04e-04 \\
            3 & 0.11 & 300 & 77 & 0.809 & 2.986 & 2.85e-04 \\
            4 & 0.17 & 298 & 78 & 0.811 & 2.922 & 4.45e-04 \\
            5 & 0.24 & 291 & 79 & 0.813 & 2.837 & 6.53e-04 \\
            6 & 0.28 & 306 & 80 & 0.815 & 2.750 & 7.60e-04 \\
            7 & 0.26 & 311 & 81 & 0.817 & 2.662 & 7.09e-04 \\
            8 & 0.33 & 307 & 82 & 0.819 & 2.565 & 8.91e-04 \\
            9 & 0.37 & 296 & 83 & 0.820 & 2.442 & 1.02e-03 \\
            10 & 0.39 & 313 & 84 & 0.822 & 2.242 & 1.10e-03 \\
            11 & 0.40 & 311 & 84 & 0.824 & 1.962 & 1.16e-03 \\
            12 & 0.40 & 308 & 84 & 0.825 & 1.651 & 1.16e-03 \\
            13 & 0.39 & 323 & 83 & 0.827 & 1.447 & 1.12e-03 \\
            14 & 0.48 & 327 & 83 & 0.829 & 1.236 & 1.40e-03 \\
            15 & 0.32 & 314 & 82 & 0.831 & 1.043 & 9.85e-04 \\
            \hline
        \end{tabular}
    \end{table}
    
    \begin{table}
        \centering
        \caption{Same as Table~\ref{tab:3MSunATON} for a 5 M$_\odot$ simulation.}
        \label{tab:5MSunATON}
        \begin{tabular}{lcccccr}
            \hline
            TDU \# & $\lambda$ & $T_\text{max}$ & $T_\text{TDU}$ & $M_\text{core}$ & $M_\text{env}$ & $M_\text{dredge}$\\
            \hline
            1 & 0.03 & 318 & 86 & 0.873 & 3.640 & 4.29e-05 \\
            2 & 0.07 & 314 & 87 & 0.874 & 3.547 & 1.09e-04 \\
            3 & 0.10 & 310 & 87 & 0.876 & 3.432 & 1.66e-04 \\
            4 & 0.14 & 306 & 88 & 0.877 & 3.297 & 2.19e-04 \\
            5 & 0.19 & 303 & 88 & 0.878 & 3.138 & 3.04e-04 \\
            6 & 0.22 & 311 & 88 & 0.880 & 2.950 & 3.62e-04 \\
            7 & 0.25 & 313 & 88 & 0.881 & 2.741 & 4.25e-04 \\
            8 & 0.28 & 313 & 89 & 0.882 & 2.509 & 4.74e-04 \\
            9 & 0.26 & 307 & 88 & 0.883 & 2.295 & 4.59e-04 \\
            10 & 0.29 & 319 & 88 & 0.885 & 2.083 & 5.08e-04 \\
            11 & 0.29 & 295 & 88 & 0.886 & 1.893 & 5.11e-04 \\
            12 & 0.30 & 306 & 88 & 0.887 & 1.713 & 5.23e-04 \\
            13 & 0.27 & 326 & 87 & 0.888 & 1.544 & 4.85e-04 \\
            14 & 0.30 & 302 & 87 & 0.890 & 1.395 & 5.24e-04 \\
            15 & 0.33 & 313 & 86 & 0.891 & 1.265 & 5.90e-04 \\
            16 & 0.22 & 309 & 86 & 0.892 & 1.143 & 4.15e-04 \\
            17 & 0.12 & 335 & 85 & 0.894 & 1.037 & 2.19e-04 \\
            18 & 0.19 & 316 & 85 & 0.895 & 0.943 & 3.33e-04 \\
            19 & 0.25 & 314 & 84 & 0.897 & 0.856 & 4.39e-04 \\
            \hline
        \end{tabular}
    \end{table}
    
    The TDU efficiency $\lambda$ parameter\footnote{$\lambda$ = $\Delta M_\text{dredge}$/$\Delta M_\text{core}$, where $\Delta M_\text{dredge}$ is the TDU penetration into the He intershell and $\Delta M_\text{core}$ is the growth of the H-exhausted core during the previous interpulse.} and the ratio of dredged mass to envelope mass are the best indicators of the efficiency of envelope enrichment of each TDU event. The former is the measure of the proportion of the \textit{s}-process enriched mass that is dredged to the surface, and the latter indicates the dilution of the \textit{s}-process elements in the convective envelope. Higher values of any of these two quantities favour the stellar surface \textit{s}-process enrichment, either by introducing more freshly-synthesised material into the envelope, or reducing the dilution factor. The $\lambda$ parameter found in the \textsc{aton} models is generally lower than that found in other evolutionary codes, with the exception of \textsc{fruity}\footnote{http://fruity.oa-teramo.inaf.it/}. For example, \citet{Karakas2014NoRev} finds $\lambda_\text{max} = 0.93$ for a 4 M$_\odot$ star at a metallicity of $Z = 0.014$, while \citet{Pignatari2016} have $\lambda_\text{max} \approx 1.1$ for the 4 M$_\odot$ star at a metallicity of $Z = 0.02$, in contrast with the \textsc{aton} $\lambda_\text{max}$ of $0.48$ for the same initial mass.
    
    Regarding the temperatures, \textsc{aton} reaches higher temperatures both in the He intershell and at the base of the convective envelope (BCE) compared to other evolutionary models of the same core mass. These differences are determined by the use of the FST model for turbulent convection, which significantly affects the thermal stratification of the stellar regions laying close to the bottom of the outer convective envelope. These higher temperatures can affect both the $^{22}$Ne and $^{13}$C neutron sources. For the $^{22}$Ne neutron source, the higher temperature in the He intershell leads to a stronger activation than in other nucleosynthesis codes. In fact, this neutron source is active in the three masses presented here, with the 3 M$_\odot$ star showing a marginal activation in those TPs where the maximum temperature reaches just above the 300 MK threshold.
    
    For the $^{13}$C neutron source, the higher BCE temperature reached in \textsc{aton} translates into an activation of the HBB at lower masses than other codes (see e.g., Sect.~7 of \citealp{Ventura2018}). In fact, the HBB already activates in \textsc{aton} around an initial mass of 3.5 M$_\odot$ (\citealp{GarciaHernandez2013} and references therein), which means that the only carbon star (a star in which the surface C/O $> 1$) of the set presented here is the 3 M$_\odot$ model. This is because the HBB activation burns C through proton captures, decreasing its surface abundance. Higher BCE temperatures also lead to more hot TDU events which, as mentioned in Sect.~\ref{sec:Sprocess}, may inhibit the $^{13}$C neutron source.
    
    \subsection{SNUPPAT}
    \label{sec:Snuppat}
    
    \subsubsection{Nuclear network solver}
    \label{sec:NucSolver}
    
    A post-processing code dedicated to the \textit{s} process must correctly follow the nucleosynthesis equations for a large number ($> 300$) of species. Because nucleosynthesis processes are mathematically described by a coupled system of ordinary differential equations, and the reaction rates vary greatly between species (forming what is commonly known as a \textit{stiff} system of equations), numerical instabilities arise during the integration unless specific care is taken in the choice of the integration method (see \citealp{Longland2014}). Common knowledge of numerical integration states that to avoid these numerical instabilities, an implicit or semi-implicit (linearized implicit) method is needed. The downside is that the number of operations necessary to advance each time step is greatly increased, which makes finding a fast and reliable method a priority. Our particular choice of numerical integration scheme is based on the Bader-Deuflhard (BD) method (\citealp{Bader1983, Deuflhard1983}), a semi-implicit method with variable order of accuracy and automatic time step selection. The basic idea revolves around the use of a semi-implicit modification of the Gragg-Bulirsch-Stoer (GBS) algorithm (\citealp{Gragg1965, Bulirsch1966}). Here we were able to combine some of the techniques from the BD algorithm with a Patankar-type algorithm (e.g., \citealp{Burchard2003, Burchard2005}) to obtain a completely original stable explicit solver for nucleosynthesis equations. We refer to this solver as the Patankar-Euler-Deuflhard (PED) solver.
    
    Before explaining the PED solver, we describe shortly the BD algorithm, because our solver is based on it.
    
    The core of the BD algorithm solves the initial values problem
    \begin{equation}
        \begin{aligned}
            y' &= f(y)\\
            y_0 &= y(t = 0),
            \label{eq:InitialValsProb}
        \end{aligned}
    \end{equation}
    by dividing the integration step $H$ in $n$ substeps of size $h = H/n$ with the following discretisation \citep{Bader1983}:
    \begin{equation}
        \begin{aligned}
            \Delta_0 &= (I - h J_{y_0})^{-1} h(\eta_0)\\
            \Delta_k &= \Delta_{k - 1} + 2(I - h J_{y_0})^{-1} [hf(\eta_k) - \Delta_{k - 1}]\\
            \Delta_n &= (I - h J_{y_0})^{-1} [hf(\eta_n) - \Delta_{n - 1}]\\
            \eta(H, h) &= \eta_n + \Delta_n,
        \end{aligned}
    \end{equation}
    where $\eta_0 = y_0$, $\eta_k = \eta_{k - 1} + \Delta_{k - 1}$, $k = 1, ..., n - 1$, $J_{y_0}$ is the initial Jacobian ($df/dy$) calculated for $y = y_0$, and $I$ is the identity matrix of the same size as the Jacobian. The array $\eta(H, h)$ is the approximate solution of Eq. (\ref{eq:InitialValsProb}) after a time $H$ for substeps of size $h$. Unfortunately, this algorithm requires the solution of the algebraic equation $(I - h J_{y_0}) Y  = X$ for different known X vectors. Although this can be made faster with the LU algorithm by decomposing once $I - h J_{y_0}$ per combination of $y_0$ and $h$, it is still the slowest operation in the method.
    
    The error expansion of this discretisation can be written as
    \begin{equation}
        \eta(H, h) = y(H) + g_1(H) h^2 + g_2(H) h^4 + ...,
    \end{equation}
    and allows for an efficient Richardson extrapolation such as that from the GBS algorithm. As an example, if the solution is re-calculated for a smaller substep of $h' = h/2$, one can see that
    \begin{equation}
        \eta(H, h/2) = y(H) + g_1(H) h^2/4 + g_2(H) h^4/8 + ...,
    \end{equation}
    with the drawback of having to perform twice the number of operations. One can then combine the solution with $h$ and $h'$ to remove the $g_1(H)$ term, improving the convergence speed of the method. In \cite{Deuflhard1983}, the preferred extrapolation method is
    \begin{equation}
        T_{i,k} = T_{i,k-1} + \frac{T_{i,k-1} - T_{i-1,k-1}}{\left(\frac{n_i}{n_{i-k+1}}\right)^2 -1},
        \label{eq:BD-extrap}
    \end{equation}
    where $T_{i,0} = \eta(H, h_i) = \eta(H, H/n_i)$. Once the $T_{i, k}$ term has been calculated, the $g_k(H)$ term has been removed from the error expansion.
    
    One can then calculate the $k$-th relative error $\epsilon_k$ by using the $T_{i, k}$ (see below for our particular choice) and test whether $\epsilon_k$ is below the desired tolerance \textsc{tol}. If the maximum $n_i$ is reached before that, then $H$ must be reduced and the calculation tried again.
    
    The next step is to choose the correct following $H$ for the calculation. The basic formula for the error $\epsilon_k$ is
    \begin{equation}
        H_k = H\left(\frac{\textsc{tol}}{\epsilon_k}\right)^\frac{1}{2k + 1}.
        \label{eq:BD-step-control}
    \end{equation}
    Although, in principle, one can choose the last $k$, such that $\epsilon_k < \textsc{tol}$, \cite{Deuflhard1983} takes into account the amount of work needed to arrive to that order, and whether it is better to aim for a smaller $k$, thus reducing the time step, or to attempt for an order increase (higher $k$ and time step). The description of these steps is complex and outside of the scope of this paper, and if desired, they can be included in the same way for the BD algorithm and our solver.
    
    The basic difference between our novel solver and the BD algorithm is the Patankar-Euler discretisation. In essence, this discretisation can be used for Eq. (\ref{eq:InitialValsProb}) if the right-hand side $f(y)$ can be written as
    \begin{equation}
        f(y) = K(y) - D(y)y,
    \end{equation}
    with $K, D > 0$, as is the case for nuclear networks. Then, Eq. (\ref{eq:InitialValsProb}) can be solved by using the following discretisation:
    \begin{equation}
        \begin{aligned}
            y_{i + 1} &= y_i + (K_i - D_i y_{i + 1})h\\
            y_{i + 1} &= \frac{y_i + hK_i}{1 + hD_i},
            \label{eq:PE-disc}
        \end{aligned}
    \end{equation}
    where $h$ is the time step chosen, $K_i = K(y_i)$, and $D_i = D(y_i)$.
    
    It is immediate to see from Eq. (\ref{eq:PE-disc}), that this discretisation is unconditionally stable for any positive $h$. That is, that $y_{i + 1}$ remains bounded even when $h$ does not. However, the method itself does not require solving any algebraic equation, because all the terms on the right-hand side are known, making it much faster than the BD algorithm.
    
    Unfortunately, the basic Patankar-Euler algorithm converges like the Euler method, that is, with a global truncation error proportional to $h$. We can accelerate the solver using a Richardson extrapolation, like for the BD algorithm. Unlike the BD algorithm, the Patankar-Euler error expansion does not contain only even powers of $h$. In fact, it can be shown to behave as
    \begin{equation}
        \eta(H, h) = y(H) + g_1(H) h + g_2(H) h^2 + g_3(H) h^3 + ...,
    \end{equation}
    It follows then that, for the Richardson extrapolation, we must use the recursive expression
    \begin{equation}
        T_{i,k} = T_{i,k-1} + \frac{T_{i,k-1} - T_{i-1,k-1}}{\frac{n_i}{n_{i-k+1}} -1},
    \end{equation}
    instead of Eq. (\ref{eq:BD-extrap}).
    
    The third and final difference between the BD algorithm and our novel solver, is that, instead of using Eq. (\ref{eq:BD-step-control}) for the step size predictor, we must use
    \begin{equation}
        H_k = H\left(\frac{\textsc{tol}}{\epsilon_k}\right)^\frac{1}{k + 1}.
    \end{equation}
    
    With these three changes, we end up with a solver that combines the speed of the explicit methods with the stability of the implicit methods. The potential limitations of this solver lay on the stiffest of problems, however, our tests so far (Yag\"ue et al., in preparation), indicate that this solver behaves well for H burning, He burning, and both the main and the weak \textit{s} process. A typical number of total substeps for the harder global integration steps during the \textit{s} process may be of the order of 1000, divided among smaller integration steps of irregular size $H_i$ each. We show the comparisons for both the main and weak \textit{s} process between the Backwards-Euler, the BD algorithm and this solver in Sect.~\ref{sec:SolverComparison}.
    
    Given that we have a powerful solver for the nucleosynthesis equations, and with the intention to easily parallelize the code, we have chosen to solve the changes to the abundance due to mixing and nucleosynthesis separately. We detail our specific numerical choices for each case below.
    
    The use of an explicit method with automatic time step selection and accuracy monitoring makes it straightforward to follow all the coupled nucleosynthesis equations from neutrons to $^{210}$Po. For the results presented in this paper, we are using the 328 element nuclear network provided by \citet{Karakas2016}. Numerically, we allow a maximum average relative error of $5 \times 10^{-5}$, this relative accuracy is calculated through the expression
    \begin{equation}
        \epsilon_k = \sqrt{\frac{1}{N} \sum_{i = 1}^N \left(\frac{T_{k,k-1}^i - T_{k,k}^i}{\max(|T_{k,k-1}^i|,y_\text{scale})}\right)^2},
        \label{eq:relError}
    \end{equation}
    where $T_{k,j}^i$ are the $j$-th order Richardson extrapolations of the solution to species $i$ for the $k$-th index in the extrapolation sequence, $y_\text{scale}$ is a numerical lower limit to the abundances ($10^{-24}$ in these calculations), $N$ is the total number of species, and we use the extrapolation error as a proxy of the numerical approximation error. We note that we prefer not to use the initial abundances $Y_i$ for scaling the error to avoid some undesirable behaviours corresponding to rapid changes in the abundances. In particular, we want to prevent an artificial bottleneck to the calculation if it turns out that some $T_{k,k}^i$ is much larger than its corresponding initial abundance. In the same vein, we would like to avoid an underestimation of the error if the $T_{k,k}^i$ end up being much lower than the initial abundances. Tests done to the numerical methods used in \textsc{snuppat} show that Eq.~(\ref{eq:relError}) is a good approximation of the error to the analytical solution for the required precision.
    
    \subsubsection{Mixing solver}
    
    We have opted for a simplistic instantaneous mixing for the convective regions. However, for the overshoot mixing needed for the formation of the $^{13}$C pocket (see discussion in Sect.~\ref{sec:Sprocess}), we have opted for a so-called linear, or advective, approach similar to that of \citet{Straniero2006}, as opposed to the diffusive approach by \citet{Herwig1997} and \citet{Goriely2004}. We couple the mixing and nucleosynthesis mechanisms by using an operator-splitting approach, which consists on solving each process separately by solving the first process alone and using its solutions as initial values for the second process\footnote{As an example, consider the linear equation $y' = (a - b)y$ with solution $y = y_0\exp[(a - b)\Delta t]$. We obtain the same expression by solving first the equation $y' = ay$ for $\Delta t$ and then taking the solution $y = y_0\exp(a\Delta t)$ as the initial value for the equation $y' = -by$, and solving it for $\Delta t$ as well. Although the operator-splitting method is only exact for linear equations, it remains a useful technique for non-linear equations too.}. The differential equations we are using for overshoot mixing are
    \begin{equation}
      \begin{aligned}
          \frac{\partial Y^c}{\partial t} &= \sum_j \tau_j \left(Y^j - Y^c\right)\\
          \frac{\partial Y^j}{\partial t} &= \xi_j\left(Y^c - Y^j\right),
          \label{eq:ovMix}
      \end{aligned}
    \end{equation}
    where $Y^c$ is the instantaneously mixed abundance in the convective zone, $Y^j$ is the abundance in the radiative shell $j$, $\xi_j$ is inverse of the time it takes for a mass element to travel from the convective zone to the shell $j$, and $\tau_j$ is related to $\xi_j$ through abundance conservation. The coefficient $\xi_j$ can be calculated as
    \begin{equation}
        \frac{1}{\xi_j} = \int_{r_c}^{r_j}\frac{dr}{u_c\left(\frac{P}{P_c}\right)^{\pm\left(\omega f_\text{thick}\right)^{-1}}} = \int_{r_c}^{r_j}\frac{dr}{u},
        \label{eq:xiInteg}
    \end{equation}
    where $r_c$ and $r_j$ are the spatial coordinates of the edge of the convective region and the overshooted shell, respectively, and $u_c$ is the turbulent velocity at the edge of the convective region. The relation between the overshoot velocity and $u_c$, as well as the description for $f_{\text{thick}}$, can be found in \citet{Ventura1998}. As usual for these kinds of prescriptions, $\omega$ is the free overshoot parameter that is to be calibrated with observations. Values of $\omega$ larger than $0.14$ are not explored due to numerical limitations. Specifically, values above $\omega \sim 0.2$ result in the introduction of protons into the PDCZ, where the instantaneous convection adopted here mixes them into the He-exhausted core, producing non-physical results.
    
\begin{figure}
        \includegraphics[width=\hsize]{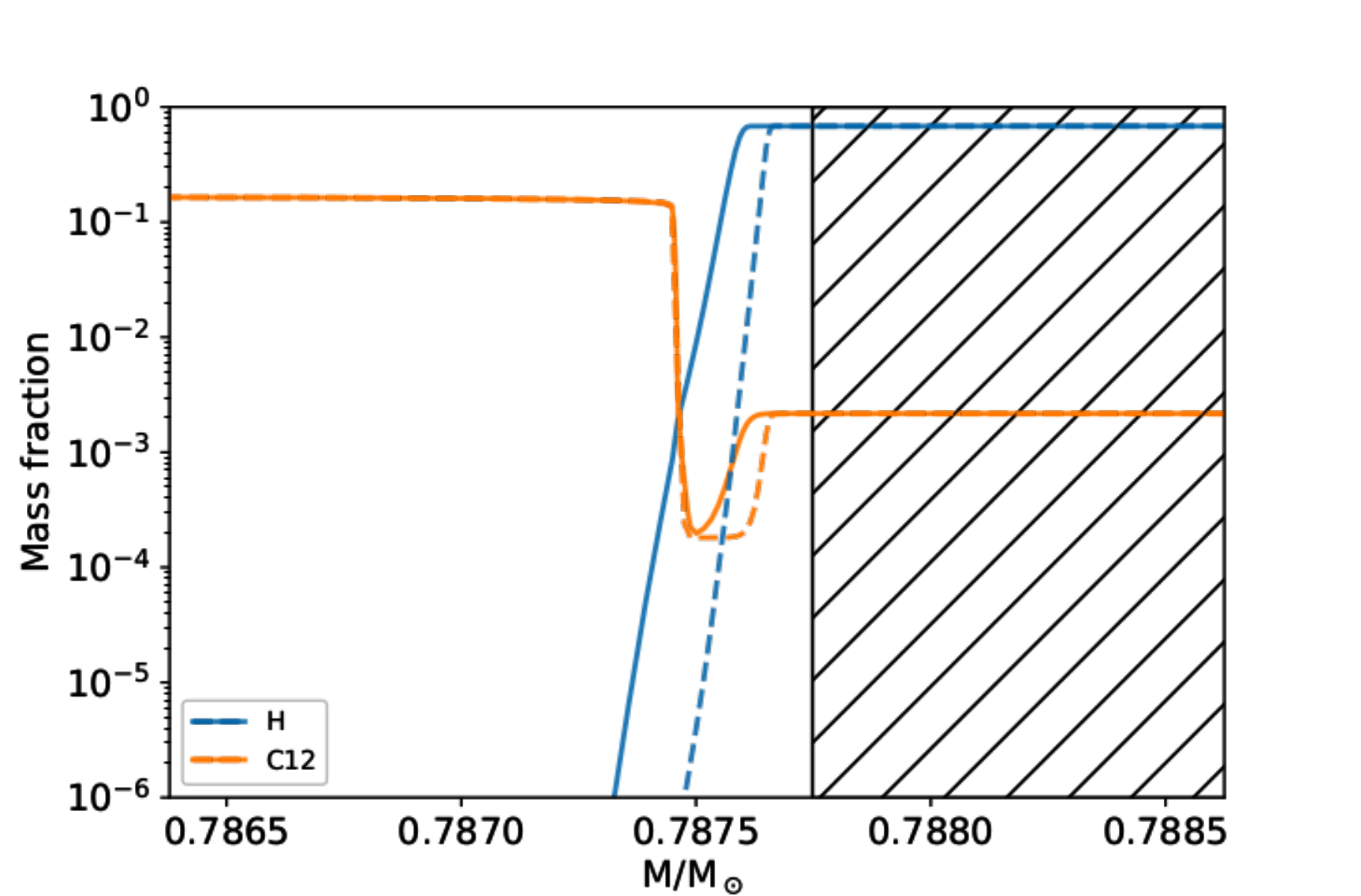}
        \caption{Abundance profiles resulting from Overshoot mixing using Eqs. (\ref{eq:ovMix}) with the parameters $\omega = 0.10$ (dashed) and $\omega = 0.14$ (solid) for a 4 M$_\odot$ case. Not only the extent of the overshooted region grows with the increase of the free parameter, but also the slope of the proton profile decreases, which results in wider effective $^{13}$C pockets. The convective envelope is shown as the hatched area.}
        \label{fig:ovMixing}
    \end{figure}

    The abundance profile resulting from solving the system (\ref{eq:ovMix}) is a straight line in a logarithmic plot, with the $\omega$ parameter controlling the extent of the mixed region (known as partial mixing zone). A specific example for a 4 M$_\odot$ model is depicted in Fig.~\ref{fig:ovMixing}, showing the difference between using $\omega = 0.10$ and $\omega = 0.14$.
    
    The fact that our overshoot method results in an exponential proton profile into the He intershell helps to compare it with other similar profiles, such as those detailed in \citet{Buntain2017}. From that work and others, we know that we can expect an effective $^{13}$C pocket with a mass roughly half of that of the partial mixing zone. The main difference with the standard H inclusion profile of \citet{Buntain2017} is that we also find an extension of the convective envelope abundances into the He intershell. As an example, we note how in Fig.~\ref{fig:ovMixing} the convective envelope abundances extend to the radiative region, and that this extension depends on the overshoot parameter. This additional completely mixed region is unavoidable for large enough overshoot parameters, given that when a deeper section of the He intershell is reached by the partial mixing, the earlier shells experience an increased mixing efficiency. This effectively represents an extension of the TDU with respect to the Tables \ref{tab:3MSunATON}, \ref{tab:4MSunATON} and \ref{tab:5MSunATON} of around $10^{-4}$ M$_\odot$. This behaviour is also reproduced in other simulations such as those computed with the \textsc{fruity} code, as can be seen in Fig.~5 of \citet{Cristallo2009}.
    
    To simulate the $^{22}$Ne neutron source, which burns convectively, we need to make sure that our operator-splitting along with instantaneous mixing approach is well suited to simulate this neutron source activation. In general terms, there are three possibilities regarding the interaction between convective mixing and the nuclear burning: i) The nuclear burning is much faster than the convective mixing, ii) the timescales for both processes are comparable, and iii) the convective mixing is much faster than the nuclear burning. Among these three cases, only the last can be accurately simulated with an instantaneous mixing approach, while the first two require the solution of the convective mixing with a time-dependent algorithm.
    
    In order to assess which of these regimes corresponds to the $^{22}$Ne($\alpha$,n)$^{25}$Mg reaction within a PDCZ, we perform a comparison between the reaction and mixing timescales. The reaction timescale can be calculated by taking the typical PDCZ temperature in the \textsc{aton} models between 4 and 6 M$_\odot$, around 350 MK. At this temperature and with a $^4$He abundance of $\sim 0.2$ mol/g, we have a timescale for the $^{22}$Ne($\alpha$,n)$^{25}$Mg reaction of
    \begin{equation}
        \tau_{^{22}\text{Ne}} \approx 10^6~\text{s}.
    \end{equation}
    Meanwhile, the mixing timescale can be approximated with the expression
    \begin{equation}
        \tau_\text{diff} \sim \frac{L}{u},
    \end{equation}
    where $u$ is the turbulent velocity and $L$ the characteristic PDCZ length. By taking $L$ to be $\sim 10^9$ cm and $u \in [10^4, 10^6]$ cm s$^{-1}$, which are typical values for these PDCZ in \textsc{aton}, we find that $\tau_\text{diff} \in [10^3, 10^5]$, or that the mixing rate is between 10 and 1000 times faster than the burning rate. Therefore, the instantaneous convective mixing approach is a good enough simulation of reality when concerning this neutron source (e.g., \citealp{Marigo2013}).
    
    We must point out, however, that our operator-splitting approach supposes that both the nuclear burning and the mixing have a time dependence that can approximate their behaviour to the real solution for arbitrarily small time steps. This supposition is critical for two reasons: i) There are individual species, such as neutrons, for which the nuclear burning timescale is always orders of magnitude faster than any mixing process, and we cannot correctly model their behaviour if mixing instantaneously. We tested that the choice between mixing the neutrons along with the other species or burning them locally where they are created, appears to have no impact in the final \textit{s}-process overabundances. This equivalence has been calculated and reported before in the literature (\citealp{Hollowell1990}). ii) The numerical approximation error cannot be assessed when one of the components acts instantaneously. This is because any way to measure this error depends on the possibility of obtaining the final results with an ever decreasing time step until the relative difference between two consecutive solutions are bounded by a tolerance imposed by the user, which is not possible to do if one of the processes is instantaneous. Given that our mixing implementation is not time-dependent, we briefly address its limitations.
    
    First, we side-step the limitation on the calculation of the numerical approximation error by not making that calculation. Instead, we ensure a short enough time step by taking every one of the outputted \textsc{aton} models in each post-processing simulation during a TP (when the time step between two models is approximately $0.1$ years), and intercalate a number of instantaneous mixing episodes between the nuclear burning reaction calculations, such that no more than $0.02$ years pass between mixing episodes. Second, we point out that the choice of an operator-splitting solution is unavoidable when taking an instantaneous mixing approach if we wish to avoid the one-shell burning approximation (that considers a constant T and $\rho$ equal to an average on the mixing zone) to the convective regions. Conversely, operator-splitting allows for a very simple and effective parallelization of the post-processing code.
    
    The last situation where the mixing and burning happen at the same time is during the HBB. In this case, the mixing timescale at the base of the convective envelope and the proton burning timescale are comparable \citep[see, e.g.][]{Ritter2018} and can be in the timescale of years to days for \textsc{aton}. The operator-splitting method is not accurate with an instantaneous mixing approximation when the burning and mixing timescales are similar (in particular for species such as Li). However, by reducing the timestep to approximately 0.01 years (around 100 hours), \textsc{snuppat} can more closely follow the \textsc{aton} HBB nucleosynthesis. Moreover, the final \textit{s}-process abundance is not significantly affected (by less than a 4\%) when the timestep is reduced to more closely simulate the HBB nucleosynthesis. This is expected due to the little effect that this timestep change has on He intershell $^{12}$C abundance (below 0.3\%) and on the envelope hydrogen abundance (below 0.03\%) in our models.

\section{Results}
\label{sec:Results}

    \subsection{Solver comparison}
    \label{sec:SolverComparison}
    
    We have implemented the PED solver in NuPPN, where an implementation of the BD and Backwards-Euler (BE) methods already exist and are used for NuGrid nucleosynthesis calculations. This allows us to compare our solver directly to two already tested methods, showcasing its accuracy and speed. We have decided to compare it in two trajectories: a He-burning trajectory in a 12 M$_\odot$ star and a $^{13}$C pocket trajectory. For each method and trajectory, the basic timesteps have been forcibly subdivided in a number of $n = \{1, 2, 4, 8, 16\}$ steps, which the BD and PED solvers automatically divide further into substeps to calculate the Richardson expansion as described in Sec.~\ref{sec:NucSolver}. Each method was run with a tolerance of $10^{-3}$. This tolerance is the maximum error allowed according to the method error monitor. For example, for the PED solver this error is given by Eq.~(\ref{eq:relError}).
    
    \begin{figure}
        \includegraphics[width=\hsize]{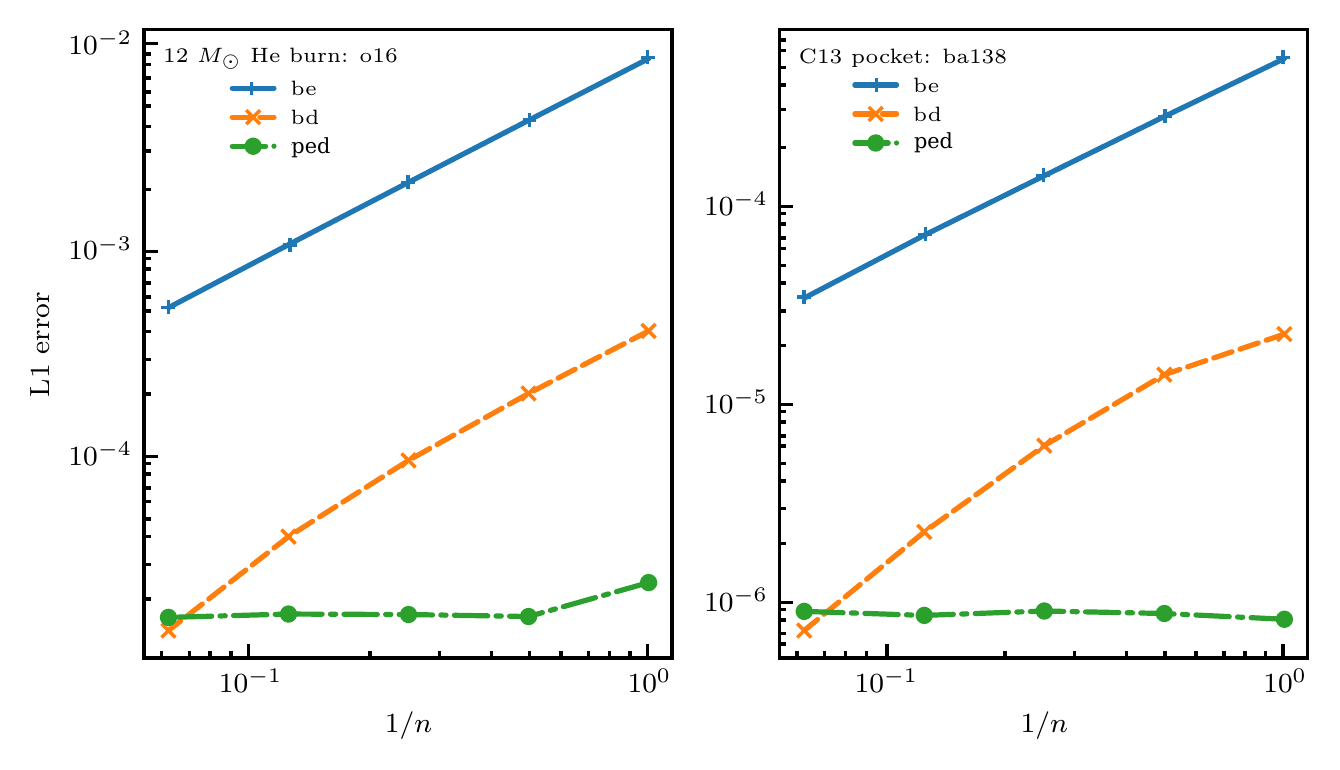}
        \caption{L1 error (sum of the absolutes of the differences) for a chosen species and process when changing the number of steps $n = \{1, 2, 4, 8, 16\}$ for each solver for the current NuPPN implementation. Each solution is compared to the solution yielded by the BD solver for $n = 32$ steps with a tolerance of $10^{-3}$. In the left panel, the final abundance for $^{16}$O in a 12 M$\odot$ He-burning trajectory. In the right panel, the final abundance for $^{138}$Ba for a $^{13}$C pocket trajectory. The BD and PED converge to the same solution. The slower convergence of the BE solver is due to its lower convergence order.}
        \label{fig:methodsErrors}
    \end{figure}
    
    In Fig.~\ref{fig:methodsErrors} we show the error of each of the methods and chosen trajectories by choosing a single relevant isotope. The solutions are compared each to a calculation with the BD solver with $n = 32$ steps. Both the BD and PED solvers converge to the same L1\footnote{Defined as the sum of the absolutes of the differences} distance to the solution by $n = 16$ steps. It is likely that the BE converges also for larger values of $n$, owing to its lower convergence order.
    
    
    \begin{figure}
        \includegraphics[width=\hsize]{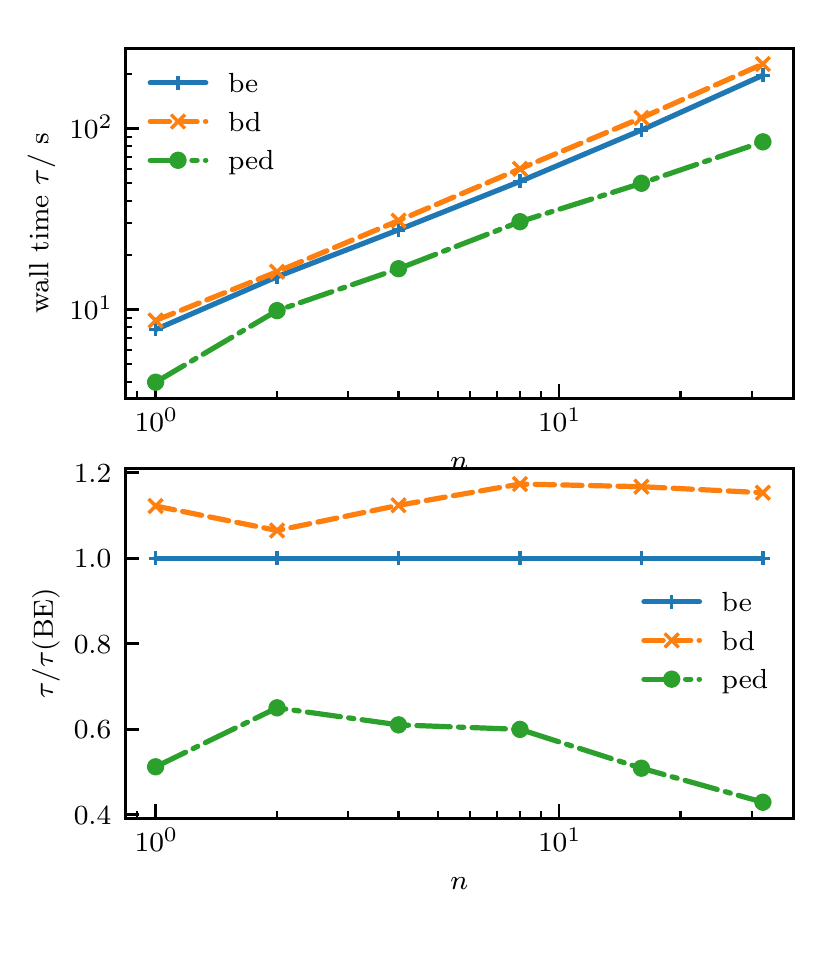}
        \caption{Comparison of the times it takes for each solver to compute a solution for a $^{13}$C pocket trajectory. In the top panel it is in absolute seconds, while in the lower panel all the times are normalised to the BE times. It is worth noting that even if BE is faster than BD, BE is less accurate for the same timesteps, as shown in Fig.~\ref{fig:methodsErrors}. The PED method is faster than the other two in all cases.}
        \label{fig:methodsWtime}
    \end{figure}
    
    In Fig.~\ref{fig:methodsWtime} we show the real time\footnote{That is, the time in seconds that the program takes to run according to a real clock} it takes for each solver to give a solution for a $n$ number of steps. In the bottom panel the times are normalised to those of the BE solver. The main difference in times between the BD and the PED solvers, despite of a similar accuracy of the solution, lies in solving the algebraic equation discussed before in the BD solver. Despite the PED solver requiring a higher number of iterations to achieve the same convergence order, the lower number of operations per iteration results in an overall much faster solution.
    
    \subsection{Intershell \texorpdfstring{$^{12}$C}{12C} abundance}
    
\begin{figure}
        \includegraphics[width=\hsize]{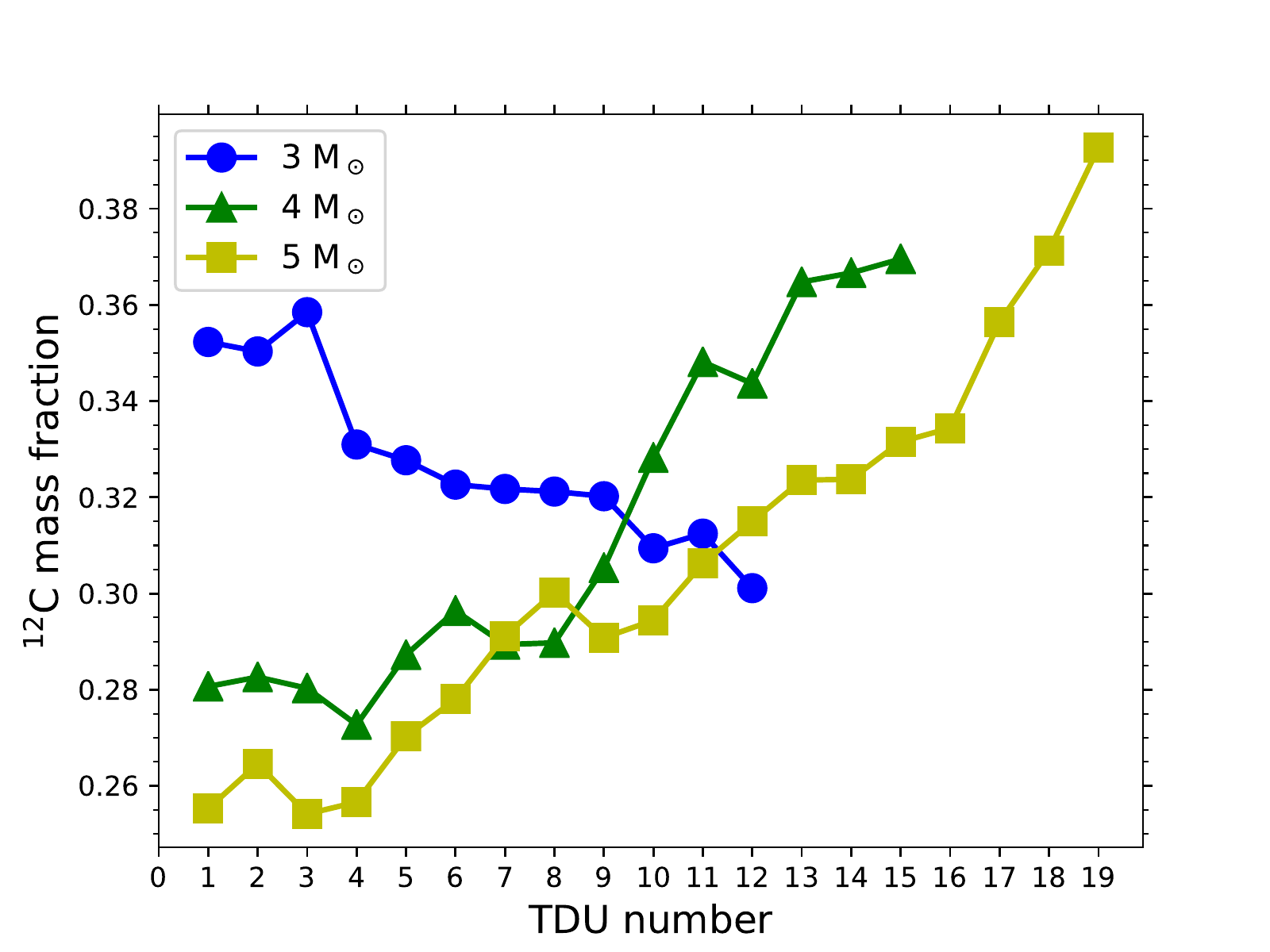}
        \caption{Evolution of He intershell $^{12}$C mass fraction during each of the TDU events represented in Tables \ref{tab:3MSunATON}, \ref{tab:4MSunATON} and \ref{tab:5MSunATON}.}
        \label{fig:c12MassFractions}
    \end{figure}

    For consistency with the \textsc{aton} models (\citealp{Ventura2018}, see also the $\zeta$ parameter in Sect.~\ref{sec:ATON}), we have included overshoot at the bottom of the PDCZ with a value for the \textsc{snuppat} overshoot parameter of $\omega = 0.002$. This overshoot affects the \textit{s}-process nucleosynthesis in several ways by increasing the $^{12}$C abundance in the He intershell, the TP luminosity and the maximum temperature, and the TDU penetration (\citealp{Herwig2000, Ventura2013}). The effects of a higher intershell temperature and TDU penetration have been discussed in Sect.~\ref{sec:Sprocess} and \ref{sec:ATON}. Consequences of an increase of the $^{12}$C abundance in the He intershell for the $^{13}$C neutron source arise because the additional $^{12}$C is turned into additional $^{13}$C. A higher $^{13}$C abundance generates more free neutrons which, therefore, increases the neutron exposure affecting the \textit{s}-process abundance distribution (e.g., see \citealp{Lugaro2003}). The values of the \textsc{aton} $^{12}$C intershell mass fraction per pulse and initial mass are shown in Fig.~\ref{fig:c12MassFractions}. The difference in the He-intershell $^{12}$C abundance evolution between the 3 M$_\odot$ and the other two models is explained by a more efficient TDU in the 3 M$_\odot$ model, which drags more $^{12}$C to the surface than is replenished by the thermal pulse.
    
    \subsection{Formation of \texorpdfstring{$^{13}$C}{13C} pocket}
    \label{sec:FormC13}
    
\begin{figure}
        \begin{center}
            \includegraphics[width=\hsize]{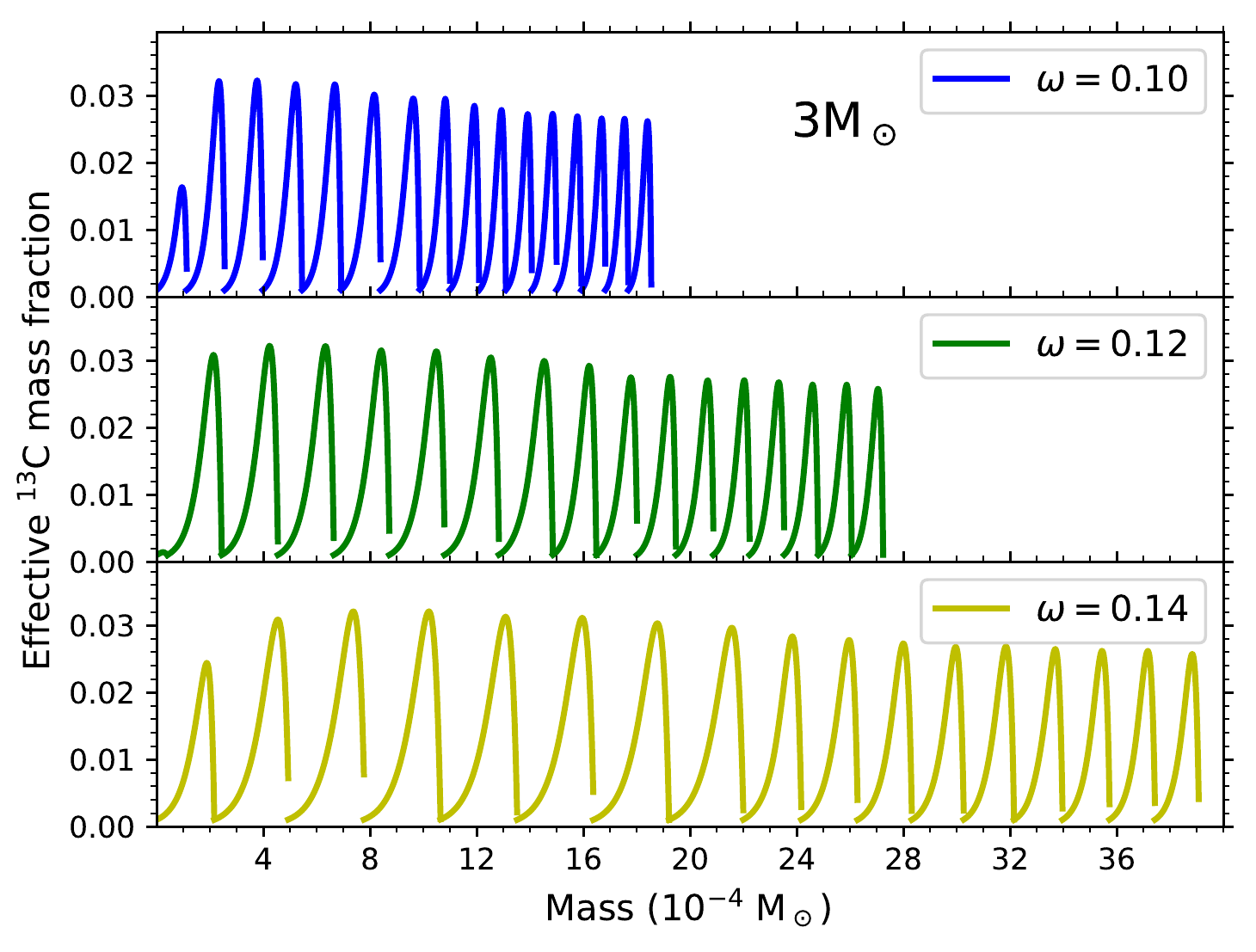}
            \includegraphics[width=\hsize]{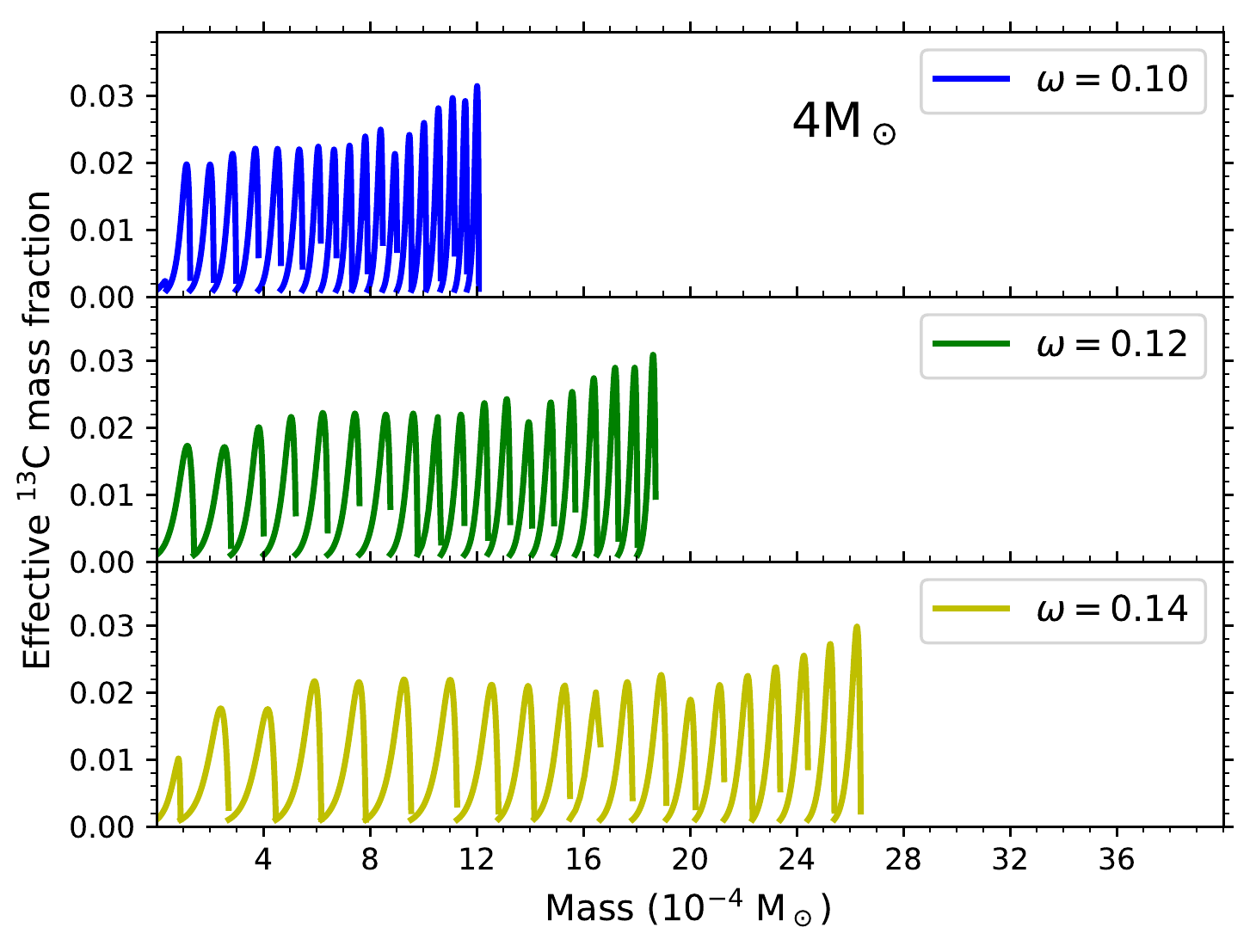}
            \includegraphics[width=\hsize]{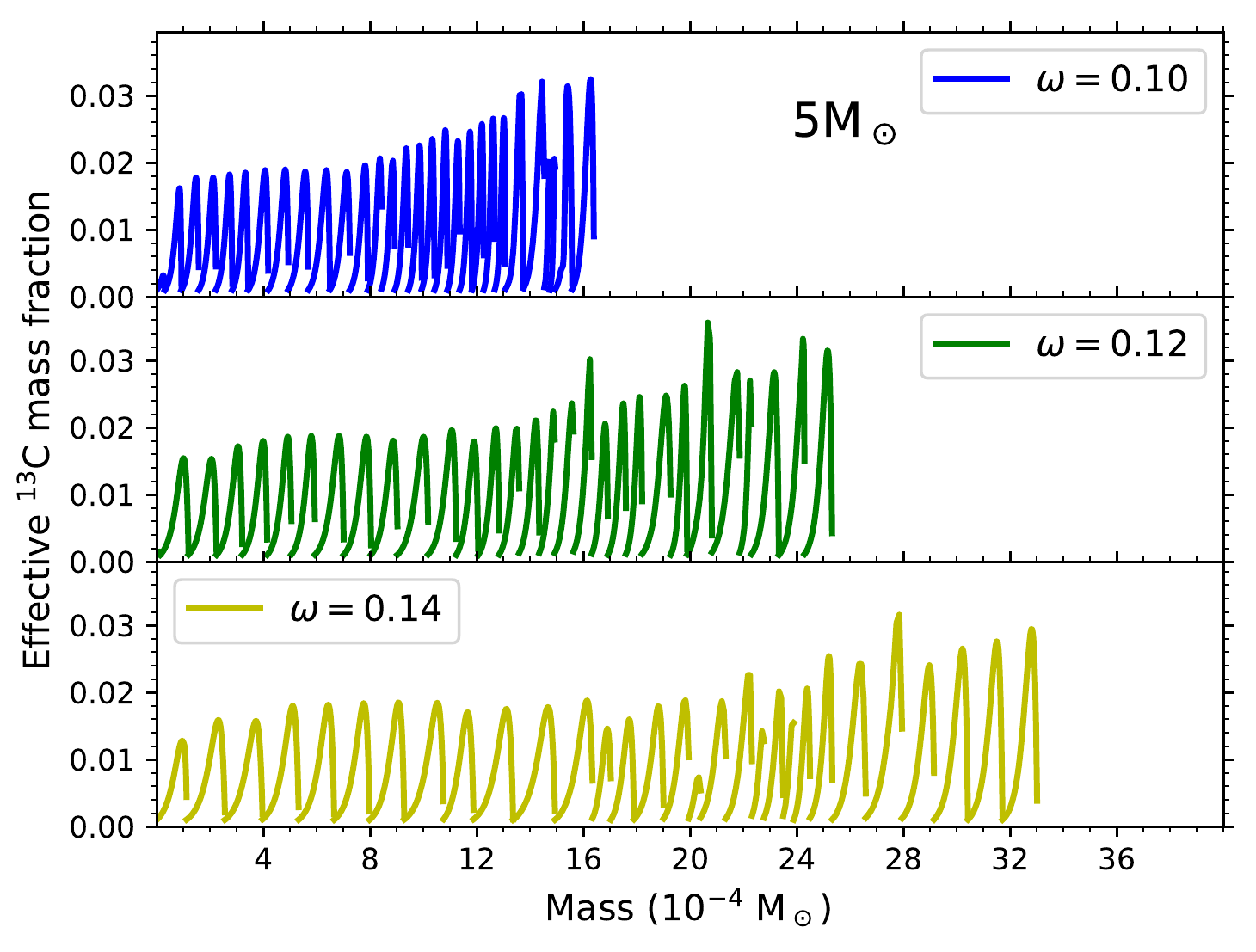}
        \end{center}
        \caption{Evolution of the effective $^{13}$C pocket in the 3, 4 and 5 M$_\odot$ stars for the overshoot parameters $\omega = 0.10$, $0.12$ and $0.14$. The y-axis shows the mass fraction of effective $^{13}$C as defined in Eq.~(\ref{eq:effPocket}), while the x-axis shows the width of the pockets in solar masses, with each pocket representing the widest mass extent of the effective $^{13}$C pocket in each interpulse. This occurs once the protons have burned completely in the He intershell, but the $^{13}$C has not started to burn yet. The pockets are artificially arranged from the earliest interpulse to the left to the latest on the right to show the temporal evolution of their shape and size. There is not a one to one correspondence of the number of $^{13}$C pockets presented here and the number of TP shown in Fig.~\ref{fig:c12MassFractions}. This is because a higher $\omega$ forces a higher TDU efficiency, forming more $^{13}$C pockets.}
        \label{fig:pockets345MSun}
    \end{figure}

    We run three separate post-processing simulations for each mass with $\omega$ = $0.10$, $0.12$, and $0.14$. The evolution and dependence of the effective $^{13}$C pocket mass extent and maximum $^{13}$C abundance on the $\omega$ overshoot parameter is shown in Fig.~\ref{fig:pockets345MSun}, and the average effective $^{13}$C pocket masses for each model are presented in Table~\ref{tab:avgC13Size}. Although the absolute TP-averaged pocket mass depends on the $\omega$ parameter; the relative maximum $X_{^{13}\text{C}_\text{eff}}$ and the mass of each pocket for a given $\omega$ is controlled by the $^{12}$C mass fraction and the proton mixing profile. Because the overshoot mixing profile (Fig.~\ref{fig:ovMixing}) depends on $\xi_j$ (that is, the inverse of the time it takes for a mass element to cover the distance from the convective envelope to the shell $j$) calculated by Eq.~(\ref{eq:xiInteg}), we conclude that the relative differences between pockets for the same $\omega$ at different interpulse periods come from differences in either the pressure gradient or the turbulent velocity $u_c$. In fact, the travelling time $1/\xi_j$ increases with a steeper pressure profile or a lower $u_c$, resulting in a smaller overshoot mixing mass, and a less massive effective $^{13}$C pocket. Steeper pressure profiles as the star evolves can be a consequence, for example, of larger $\lambda$ parameters.
    
    We point out that when running these post-processing simulations with different $\omega$ parameters, we are using the same \textsc{aton} model calculated with a $\zeta = 0.002$. This results in an inconsistency where the stellar structure does not react to the mixing of envelope material into the He intershell. This inconsistency is also found in other codes such as Monash \citep[e.g][]{Buntain2017}, and its existence means that, in these codes, the \textit{s}-process results must be understood with the caveat that there is no feedback between the overshoot (or partial-mixing zone, in the case of Monash) and the stellar structure. A more in-depth study on the feedback between overshoot extent and \textit{s}-process nucleosynthesis due to changes in stellar structure in \textsc{aton} is left to future work.
    
    \begin{table}
        \centering
        \caption{TP average of effective $^{13}$C pocket mass extent $\langle\text{M}_{\text{Eff}^{13}\text{C}}\rangle$ for each one of the models presented here.}
        \label{tab:avgC13Size}
        \begin{tabular}{lcr}
            \hline
            Mass (M$_\odot$) & $\omega$ & $\langle\text{M}_{\text{Eff}^{13}\text{C}}\rangle$ (M$_\odot$)\\
            \hline
            3 & 0.10 & 1.16$\times$10$^{-4}$ \\
              & 0.12 & 1.60$\times$10$^{-4}$ \\
              & 0.14 & 2.30$\times$10$^{-4}$ \\
            4 & 0.10 & 6.36$\times$10$^{-5}$ \\
              & 0.12 & 9.85$\times$10$^{-5}$ \\
              & 0.14 & 1.32$\times$10$^{-4}$ \\
            5 & 0.10 & 5.65$\times$10$^{-5}$ \\
              & 0.12 & 8.44$\times$10$^{-5}$ \\
              & 0.14 & 1.10$\times$10$^{-4}$ \\
            \hline
        \end{tabular}
    \end{table}
    
    The average mass extent of the pocket decreases with the initial mass due to the shrinking of the He-intershell as the initial core mass increases and as the star evolves, and the increasing in absolute value of the pressure gradient (as already described by \citealp{Cristallo2009, Cristallo2011, Cristallo2015}).
    
    Finally, the choice of overshooting scheme can have a profound effect on the formation of the $^{13}$C pocket. In particular, the inhibition of the effective $^{13}$C pocket during hot TDU events, as described by \cite{Goriely2004}, does not happen for our choice of overshooting scheme (given by Eq.~\ref{eq:ovMix}). The reason is that the overshooted proton abundances are not connected with those of the neighbouring regions, only with the convective envelope. Therefore, the protons travelling from the envelope to the deeper layers cannot burn before reaching them, even if the temperature is high enough. Although the protons may quickly burn on arrival, the profile is not affected by these temperatures. To show this, we have included a diffusive overshoot scheme in our code following \citet{Herwig1997} and performed two calculations during the same hot TDU in the 4 M$_\odot$ model where the only difference between the calculations is the scheme used for overshoot. This test shows that for the same pulse the advective overshoot does not inhibit the effective $^{13}$C pocket (see Fig.~\ref{fig:4Msun_adv}), while the diffusive one does (see Fig.~\ref{fig:4Msun_diff}). We have included the figures for this test in Appendix~\ref{sec:appInhibition}.
    
    \subsection{Stellar surface \textit{s}-process results}
    \label{sec:SurfaceResults}
    
\begin{figure}
        \begin{center}
            \includegraphics[trim = 0mm 2mm 5mm 12mm, clip, height=6.45cm]{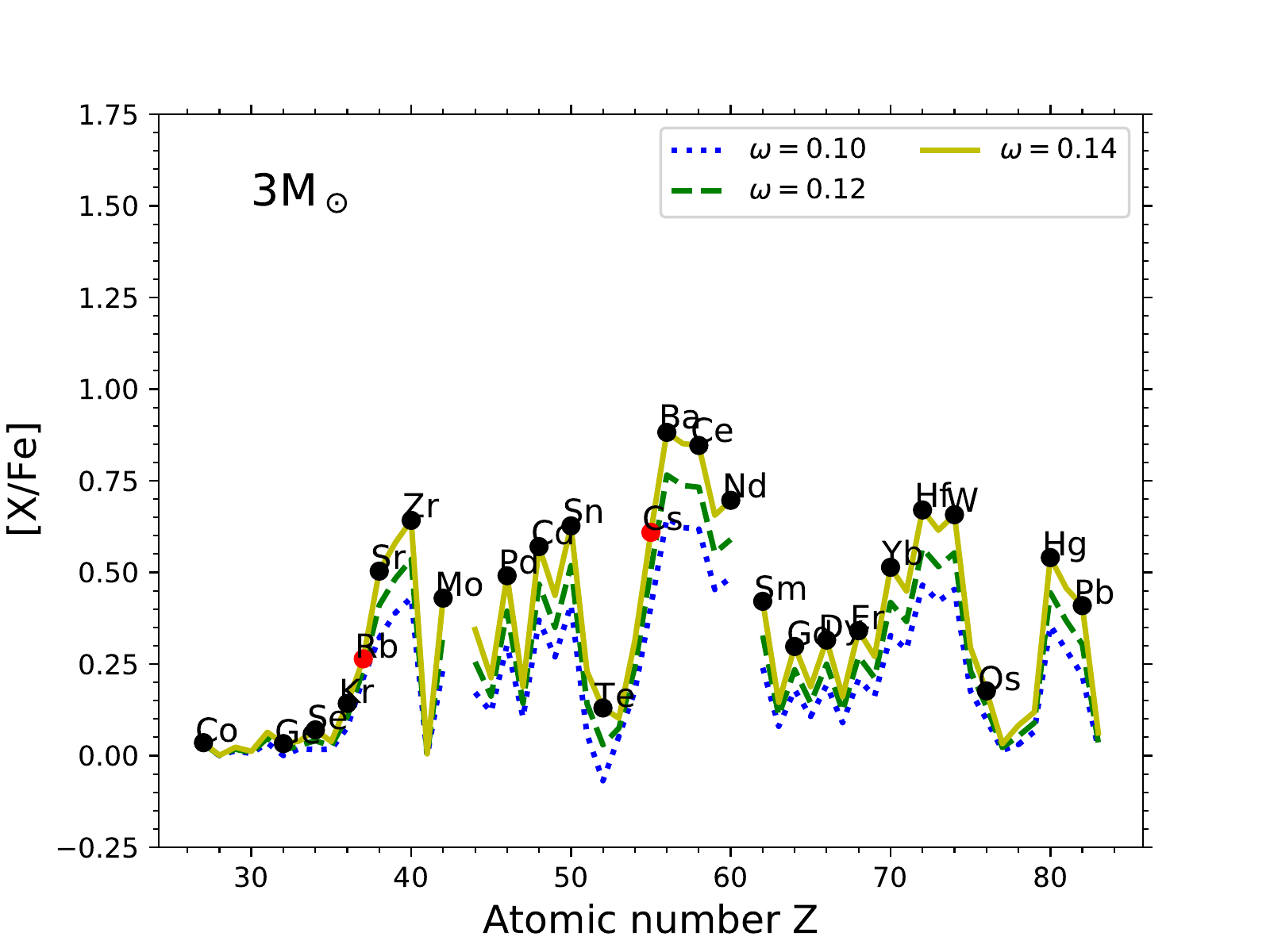}
            \includegraphics[trim = 0mm 2mm 5mm 12mm, clip, height=6.45cm]{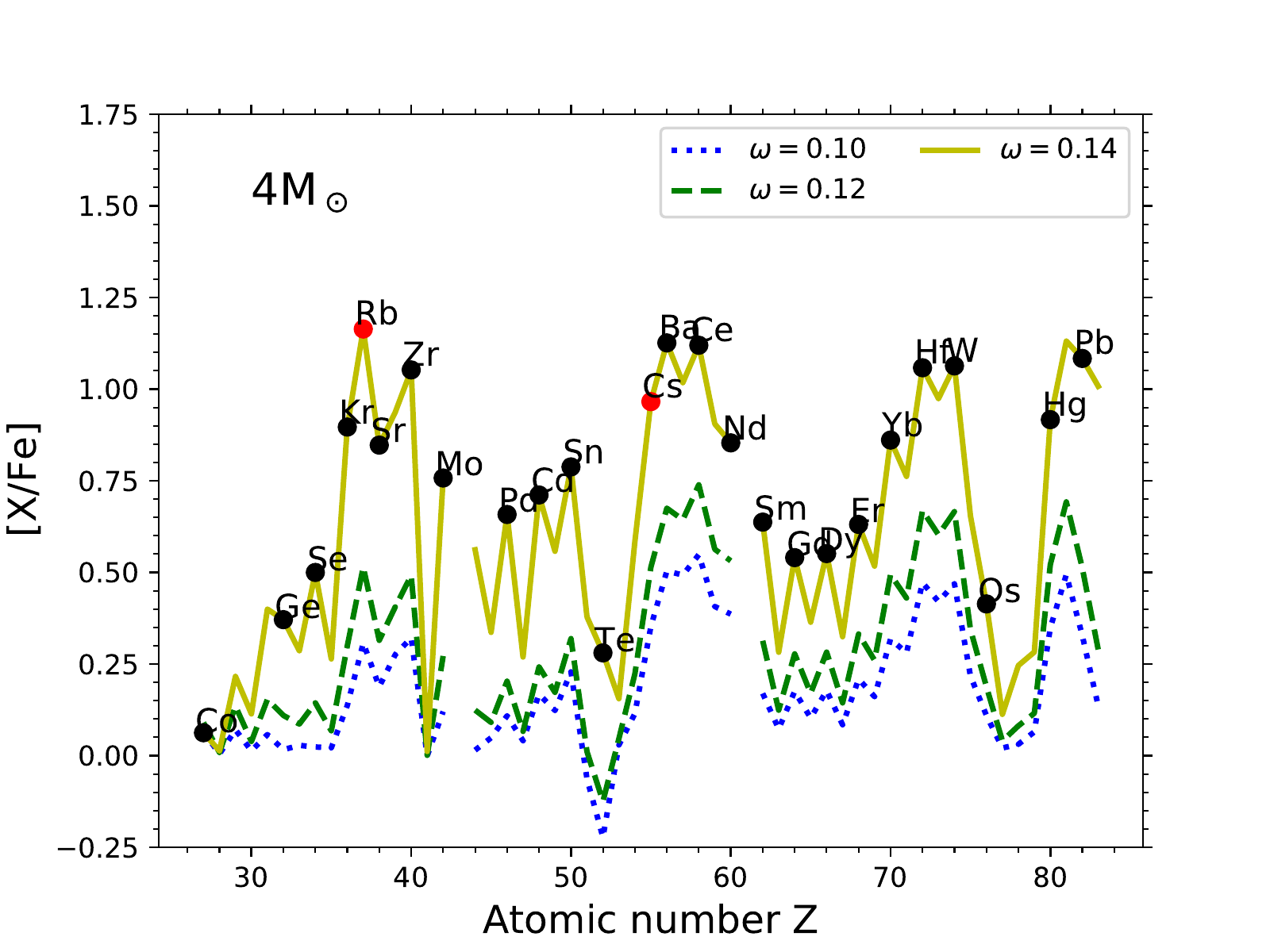}
            \includegraphics[trim = 0mm 2mm 5mm 12mm, clip, height=6.45cm]{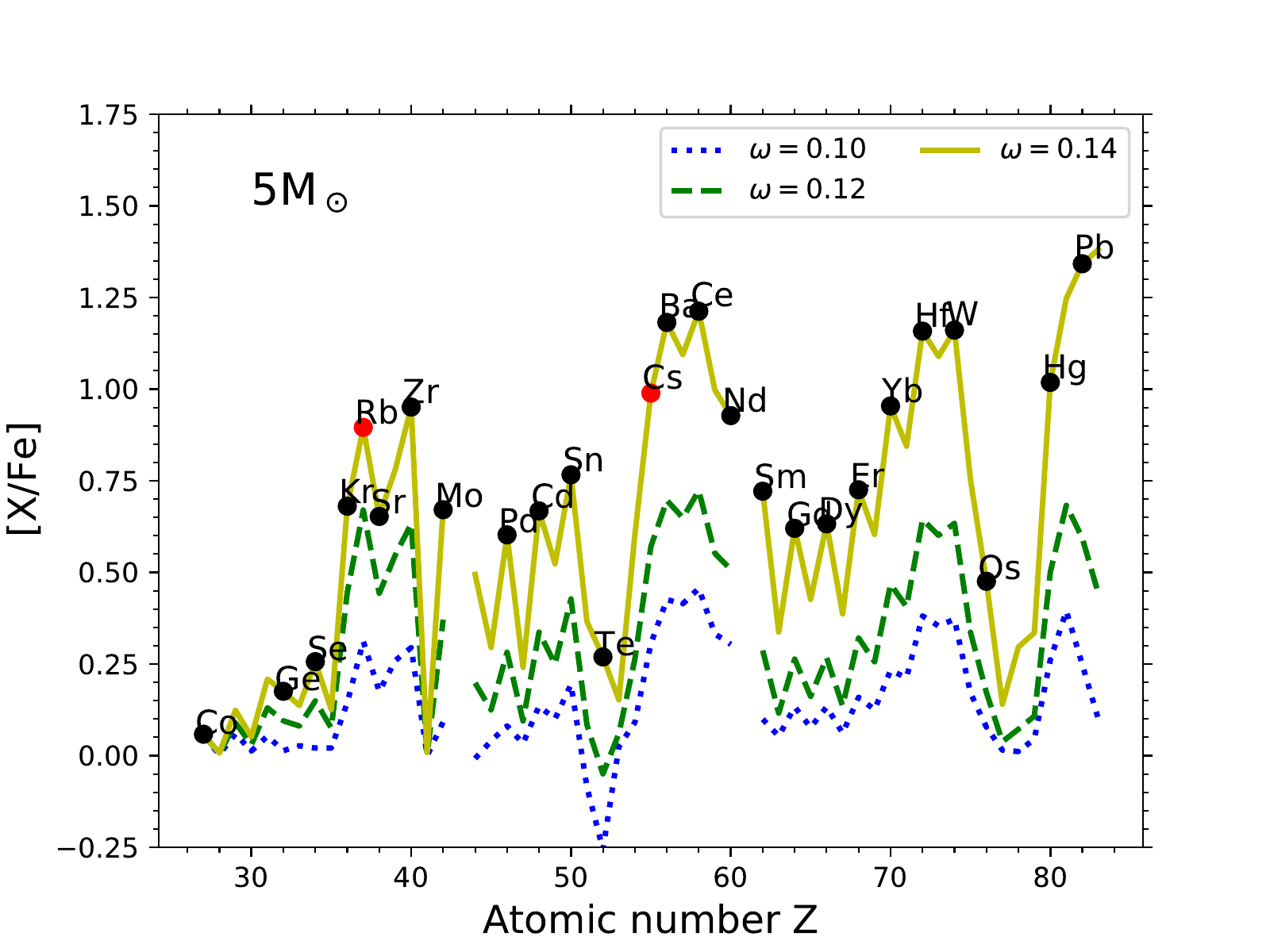}
        \end{center}
        \caption{Final stellar surface abundances of the elements heavier than Fe for the 3, 4, and 5 M$_\odot$ \textsc{aton} stars (top, middle, and bottom panel, respectively) post-processed with \textsc{snuppat} with 3 different convective overshoot parameters $\omega$. The elements marked in red are those most sensitive to the neutron density. The breaks in the abundance distribution are due to the unstable isotopes Tc (Z = 43) and Pm (Z = 61).}
        \label{fig:345MSunResults}
    \end{figure}

    The \textit{s}-process nucleosynthesis results of our models are presented in Fig.~\ref{fig:345MSunResults} and Table~\ref{tab:345MSunAbunds}.
    From Sect.~\ref{sec:Snuppat} and Fig.~\ref{fig:ovMixing} we know that the change in the overshoot parameter affects two important aspects of the nucleosynthesis calculation: the dredge-up efficiency parameter $\lambda$ and the mass extent of the effective $^{13}$C pocket. An increase in either of these two quantities enhances the overall final stellar surface overabundance values. A larger dredged-up mass mixes more material from the intershell region into the convective envelope, naturally boosting the stellar surface overabundance values. A more massive $^{13}$C pocket results in \textit{s}-process nucleosynthesis in a larger number of mass shells, which results in higher intershell overabundances and, therefore, higher stellar surface overabundances.
    
    Another critical aspect is the neutron exposure. This is tied to the $^{13}$C intershell mass fraction, which, in turn, depends on the $^{12}$C intershell mass fraction. Models with a higher $^{12}$C intershell abundance produce higher [hs/ls]\footnote{Where [hs/ls] is defined as the difference between [hs/Fe] and [ls/Fe]. The ratio [ls/Fe] is the average of [Sr/Fe], [Y/Fe], and [Zr/Fe], and the ratio [hs/Fe], is the average of [Ba/Fe], [La/Fe], and [Ce/Fe], see, e.g., \citet{Karakas2014}.\label{fn:hsLs}} ratios, that is, the relative production of the second to the first \textit{s}-process peaks (\citealp{Lugaro2003}).
    
    \begin{table}
        \centering
        \caption{Selected final stellar surface overabundances for the 3, 4 and 5 M$_\odot$ models presented in Fig.~\ref{fig:345MSunResults}. The [hs/ls] ratio is defined in Footnote \ref{fn:hsLs}.}
        \label{tab:345MSunAbunds}
        \begin{tabular}{lcccccr}
            \hline
            $\omega$ & [Rb/Fe] & [Sr/Fe] & [Ba/Fe] & [Pb/Fe] & [hs/ls] & [Pb/hs]\\
            \hline
            \multicolumn{7}{c} {3 M$_\odot$}\\
            \hline
            0.10 & 0.21 & 0.32 & 0.65 & 0.22 & 0.25 & -0.41 \\
            0.12 & 0.24 & 0.41 & 0.77 & 0.30 & 0.27 & -0.45 \\
            0.14 & 0.26 & 0.50 & 0.88 & 0.41 & 0.29 & -0.45 \\
            \hline
            \multicolumn{7}{c} {4 M$_\odot$}\\
            \hline
            0.10 & 0.31 & 0.19 & 0.50 & 0.33 & 0.25 & -0.18 \\
            0.12 & 0.51 & 0.31 & 0.68 & 0.51 & 0.28 & -0.18 \\
            0.14 & 1.16 & 0.85 & 1.13 & 1.08 & 0.14 & 0.01 \\
            \hline
            \multicolumn{7}{c} {5 M$_\odot$}\\
            \hline
            0.10 & 0.31 & 0.18 & 0.43 & 0.25 & 0.19 & -0.18 \\
            0.12 & 0.67 & 0.44 & 0.70 & 0.59 & 0.15 & -0.10 \\
            0.14 & 0.90 & 0.65 & 1.18 & 1.34 & 0.37 & 0.18 \\
            \hline
        \end{tabular}
    \end{table}
    
    We start discussing the results with the 3 M$_\odot$ star. As expected, the higher the $\omega$ parameter, the higher the stellar surface overabundances (see top panel of Fig.~\ref{fig:345MSunResults} and top rows of Table~\ref{tab:345MSunAbunds} for the 3 M$_\odot$ model). The largest differences can be found around the \textit{s}-process peaks. For example, the Ba abundance (represented by [Ba/Fe]), increases from approximately $0.7$ to $0.9$ dex when increasing $\omega$ from $0.10$ to $0.14$.
    
    The abundance distribution (that is, the relative differences between the abundances of different elements) is for the most part independent of $\omega$, with the only exception being the isotopes contributed by the $^{22}$Ne neutrons source, as can be seen from the [hs/ls] ratio that remains at $\sim 0.30$ for almost every $\omega$. The reason is that a larger $\omega$ only results in a more massive $^{13}$C pocket and greater TDU efficiency, not in a larger local neutron exposure. This independence of the abundance distribution for most isotopes with the mass extent of the partially mixed region is an intrinsic property of the models and is found also using the artificial partial mixing zone of \citet{Buntain2017}. The [Rb/Sr] ratio in this 3 M$_\odot$ model is negative, and decreases from $-0.11$ to $-0.24$ when increasing the $\omega$ parameter. This behaviour is consistent with the fact that the $^{22}$Ne neutron source is weakly activated and it is only mildly dependent on $\omega$, as more massive effective $^{13}$C pockets increase the [Sr/Fe] without affecting [Rb/Fe].
    
    In the 4 M$_\odot$ star, the largest changes in abundances when changing $\omega$ are concentrated around the \textit{s}-process peaks. However, the effect of the stronger activation of the $^{22}$Ne neutron source in this model, relative to the 3 M$_\odot$ case, is clearly seen at both the first and second peaks in the increase of the [Rb/Sr] and [Cs/Ba] ratios. This is due to the activation of branching points, as discussed in Sect.~\ref{sec:Sprocess}. The [hs/ls] ratio is positive, but lower than the 3 M$_\odot$ star. This is somewhat expected, given that the $^{22}$Ne neutron source favours the production of the first \textit{s}-process peak at this metallicity. The increase in the [Rb/Zr] ratio from $-0.01$ to $0.11$ and the [Cs/Ba] ratio before decay\footnote{The values presented in the tables and figures of this paper are decayed for $10^4$ years to remove some of the transient features displayed by short-lived isotopes such as those from $^{135}$Cs.} from $-0.04$ to $0.18$ when incrementing $\omega$ from $0.10$ to $0.14$ (not shown in Table~\ref{tab:345MSunAbunds}) is probably due to an increased amount of $^{14}$N, product of the more extended proton-mixed region in the He intershell. The additional $^{14}$N created by the larger overshoot is transmuted into $^{22}$Ne, which in turn boosts the efficiency of this neutron source.
    
    The 5 M$_\odot$ star behaves similarly to the 4 M$_\odot$ case. An increase in the $\omega$ free parameter not only boosts the total stellar surface abundances, but also changes the elements Rb and Cs that are produced by branching points that depend on the activation of the $^{22}$Ne neutron source. Critically, the $^{13}$C neutron source is still active in this star, which may be in contradiction with the few observations of solar metallicity massive AGB stars at the beginning of the TP phase available in the literature, which show the lack of Tc suggesting the non-activation of the $^{13}$C neutron source (\citealp{GarciaHernandez2013}). This is a consequence of our numerical approach to the overshoot given that, as described in Sect.~\ref{sec:FormC13}, a diffusive approach such as that of \citet{Goriely2004} can prevent the formation of the effective $^{13}$C pocket.

\section{Discussion}
\label{sec:Discussion}
    
    \subsection{Comparison with the results from other codes}
    \label{sec:CompCodes}
    
    One of the most important uncertainties of the \textit{s}-process nucleosynthesis simulations for low- and intermediate-mass stars is the formation of the $^{13}$C pocket. We have described the formation of the $^{13}$C pocket in \textsc{snuppat} in Sect.~\ref{sec:FormC13}, here we analyse the differences with the pocket formation in other simulations.
    
    The $X_{^{13}\text{C}_\text{eff}}$ profiles of the Monash models (\citealp{Buntain2017}) have the most similar shape to those obtained in this work. Those are produced by an artificial proton profile within a partial-mixing zone of mass $M_\text{mix}$ (a free parameter) at the deepest penetration of the convective envelope in the He-intershell during each TDU event. Taking the standard \textit{linear} (in a log scale) hydrogen profile from that work, which covers $\log(X_\text{H})$ from 0 to $-4$ and forms an effective $^{13}$C pocket below around $\log(X_\text{H}) = -2$, the mass extension of the effective $^{13}$C pocket is approximately half of the $M_\text{mix}$ parameter. In the case of 3 and 4 M$_\odot$, this means a $^{13}$C pocket mass extension between $5 \times 10^{-4}$ and $10^{-3}$ M$_\odot$ in their \textit{standard} models. These effective $^{13}$C pockets are between 2 and 5 times larger than the largest average presented in Table~\ref{tab:avgC13Size}.
    
    We treat overshoot in a similar fashion as the \textsc{fruity} models. However, the proton profiles presented by \citet{Cristallo2009} for a 2 M$_\odot$, Z = $0.014$ case are $\sim 3$ times larger in mass than those presented here for the 3 M$_\odot$ case (our least massive case), when considering their $\beta$ parameter (which is analogous to the parameter $\omega$ discussed in this work) in the range $0.075$ to $0.125$. This difference is likely related to the lower initial stellar mass of the \textsc{fruity} model we are comparing to, as the mass extent of the $^{13}$C pocket increases with decreasing initial mass. Another difference between the proton profiles in Figs. 1 and 2 of \citet{Cristallo2009} and in our Fig.~\ref{fig:ovMixing} is that our H profile is approximately a straight line in logarithmic representation, while in \textsc{fruity} the slope of this H profile decreases in absolute value towards the inside of the star. We attribute this difference to the specific choice of the function of the velocity decay from the border of the convective zone. In the \textsc{fruity} models this depends on the distance from the base of the convective envelope measured in $H_p$. In our models, it depends instead on the normalised pressure (see Eq.~\ref{eq:xiInteg}). The effective $^{13}$C pocket mass extent of the \textsc{fruity} models with $\beta = 0.1$ is approximately 3 times larger than what we obtain with $\omega = 0.14$.
    
    Finally, we compare to the $^{13}$C pockets presented in \citet{Battino2016} and \citet{Pignatari2016} (that is, the NuGrid models), which come from diffusive overshoot\footnote{In the NuGrid papers, the term ``convective boundary mixing'' is used instead of overshoot. However, for consistency with the nomenclature of this work, we use the term overshoot to refer to this boundary mixing as well.} as opposed the linear approach used by us and \textsc{fruity}. Despite this essential difference in the overshoot treatment, the proton profile obtained in \citet{Battino2016} is remarkably similar to what we obtain and, as a consequence, it produces an effective $^{13}$C pocket that resembles ours. In particular, from their Fig.~4 we recognise a similar shape and a mass extent of the same order of magnitude ($\sim 10^{-4}$ M$_\odot$) as those we obtain for an overshoot parameter of $\omega = 0.14$. The effective $^{13}$C pockets presented by \citet{Pignatari2016} are much smaller than what we obtain. For example, in their Fig.~9, they present an effective $^{13}$C pocket of $5 \times 10^{-6}$ M$_\odot$ for a 5 M$_\odot$ star, which is at least one order of magnitude lower than what we find for the same mass and metallicity. The difference between \citet{Battino2016} and \citet{Pignatari2016} is that in \citet{Battino2016} two diffusive steps are used. The consequence is a more extended partially mixed zone, resulting in a more massive effective $^{13}$C pocket and higher surface overabundances. For example, for a 2 M$_\odot$ model at solar metallicity, \citet{Pignatari2016} produce pockets of $\sim 5 \times 10^{-5}$ M$_\odot$, almost an order of magnitude lower than those at 3 M$_\odot$ while \citet{Battino2016} produce a $\sim 10 ^{-4}$ M$_\odot$ effective $^{13}$C pocket for a 3 M$_\odot$ simulation, which is only $\sim 2.5$ times smaller than the average mass extension of the pockets we obtain. We do not include the models from \citet{Battino2019} in the interest of clarity, because for the relevant case (3 M$_\odot$ and Z = $0.02$) the overall production sits between that of \citet{Battino2016} and \citet{Pignatari2016} and so is covered by these two extremes.
    
    \subsection{Comparison of the \textit{s}-process distributions}
    
    \subsubsection{3 M$_\odot$ models}
    
    \begin{figure*}
        \includegraphics[width=\textwidth]{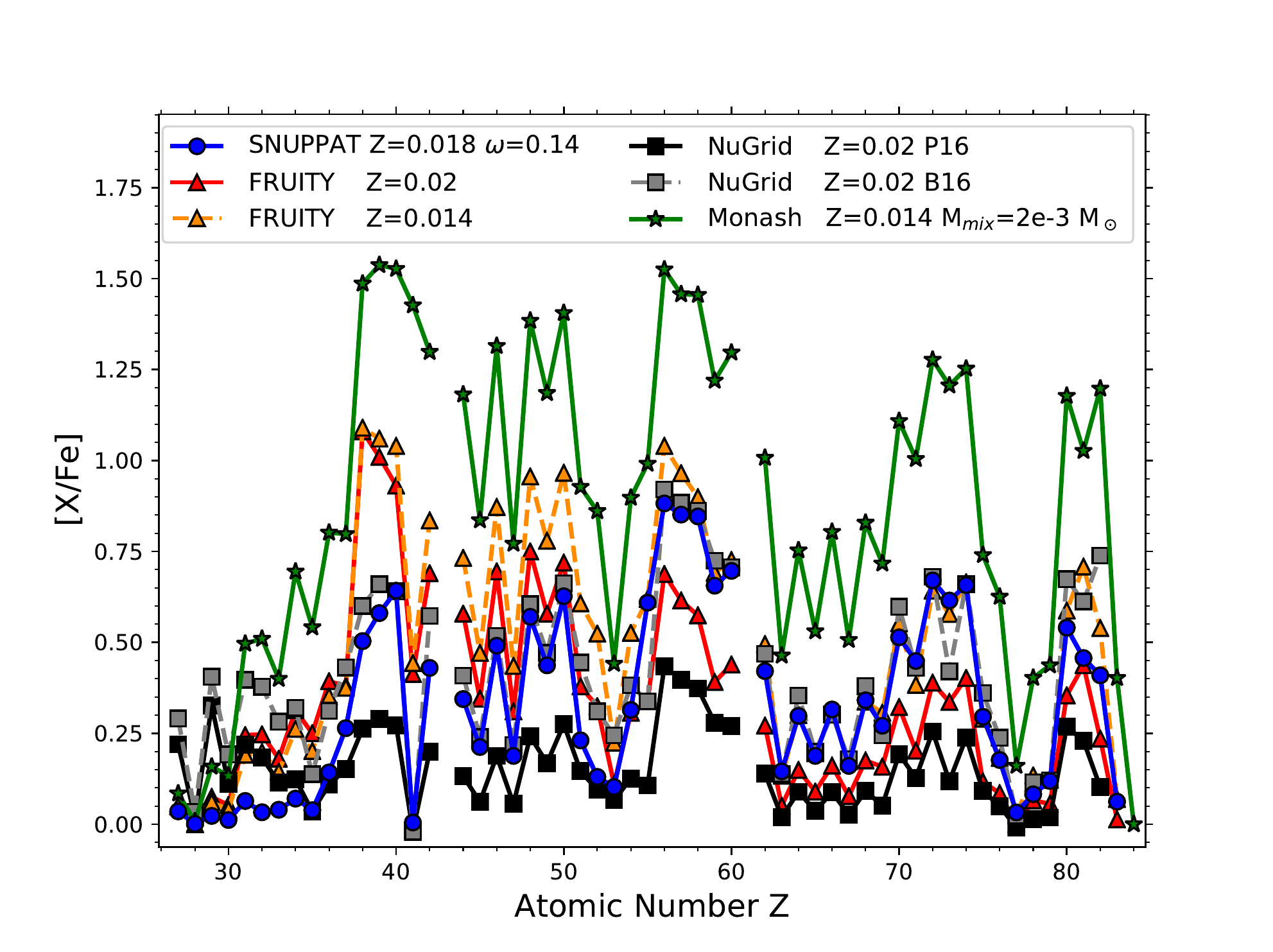}
        \caption{Comparison of \textit{s}-process nucleosynthesis results in the stellar surface of 3 M$_\odot$ models at solar metallicity of Monash, \textsc{fruity}, NuGrid, and \textsc{snuppat}. From Monash, we have selected the Z $= 0.14$ model with M$_\text{mix}$ equal to $2 \times 10^{-3}$ M$_\odot$ (\protect\citealp{Karakas2014NoRev, Karakas2016}). From the \textsc{fruity} models we have selected the Z $= 0.014$ and Z $= 0.02$ models. From NuGrid we have chosen the Z $= 0.02$ models from \protect\citet{Battino2016}, marked in the legend as ``B16'', and \protect\citet{Pignatari2016}, marked in the legend with ``P16''.}
        \label{fig:compare3MSun}
    \end{figure*}
    
    The 3 M$_\odot$ comparison is presented in Fig.~\ref{fig:compare3MSun} and shows that the Monash model produces more \textit{s}-process elements than the other codes. For example, the Monash [Ba/Fe] is $\sim 3$ times higher than what we find for an overshoot parameter of $\omega = 0.14$, while the [Sr/Fe] is 10 times higher than in our model. This can be explained by a combination of the $\lambda$ parameter and the mass extent of the effective $^{13}$C pockets as discussed in Sect.~\ref{sec:SurfaceResults}. The Monash effective $^{13}$C pocket mass extent is around $10^{-3}$ for the model selected in Fig.~\ref{fig:compare3MSun}, while in our models is $\lesssim 2.5\times10^{-4}$ M$_\odot$. The lower [hs/ls] fraction in Monash shown here is due to the lower effective $^{13}$C pocket mass fraction.
    
    Like with the Monash model, the \textsc{fruity} models show a more massive $^{13}$C pocket, but with a lower effective $^{13}$C mass fraction when compared to ours, and a similar $\lambda$ parameter. This means that a lower neutron exposure is spread over a larger number of mass shells, resulting in an overall lower [hs/ls] ratio for a comparable or even larger overabundance than in the 3 M$_\odot$ model with $\omega = 0.14$ presented here.
    
    For the NuGrid models, we have included the results from applying the two different kinds of overshoot explained in Sect.~\ref{sec:CompCodes}. \citet{Battino2016} obtain a less massive effective $^{13}$C pocket, but a larger $\lambda$ parameter, around two times larger than ours. The resulting stellar surface overabundances reported in \citet{Battino2016} are very similar to those produced by our $\omega = 0.14$ model in both \textit{s}-process peaks.
    
    As can be seen in Fig.~\ref{fig:compare3MSun}, the [hs/ls] ratio value covers a wide range in the results considered here. For example, both in the Monash models and the \textsc{fruity} models, this ratio is lower than $0.1$ dex at solar metallicity (Z $= 0.014$ in \textsc{fruity} and Monash). On the other hand, the NuGrid models (\citealp{Pignatari2016, Battino2016}) find values closer to ours for 3 M$_\odot$ models at solar metallicity (Z = $0.02$), with $\sim 0.15$ dex in \citet{Pignatari2016} and up to $0.25$ dex in \citet{Battino2016} while, when using their definition for [hs/ls]\footnote{The definition of [hs/Fe] used in \citet{Pignatari2016} is slightly different to what we use as they chose the average of [Ba/Fe], [La/Fe], [Nd/Fe], and [Sm/Fe].}, our maximum sits at $0.14$ dex.
    
    These differences are explained by the intershell abundance of $^{12}$C, which affects the $^{13}$C mass fraction generated by proton captures, as outlined in Sect. \ref{sec:ATON}.
    In the Monash models (\citealp{Buntain2017}), the intershell $^{12}$C mass fraction is below approximately $0.25$, which means that the effective $^{13}$C mass fraction remains below $0.025$, providing an [hs/ls] ratio of approximately zero. In the \textsc{fruity} models (see \citealp{Cristallo2011, Cristallo2015}), the $^{13}$C mass fraction is below $0.02$, which results in a negative [hs/ls] ratio. Finally, the NuGrid models show an intershell $^{12}$C mass fraction between $0.4$ and $0.5$ (see Fig.~10 of \citet{Battino2016}), resulting in a final [hs/ls] ratio similar to ours.
    
    \subsubsection{4 M$_\odot$ models}
    
    \begin{figure*}
        \includegraphics[width=\textwidth]{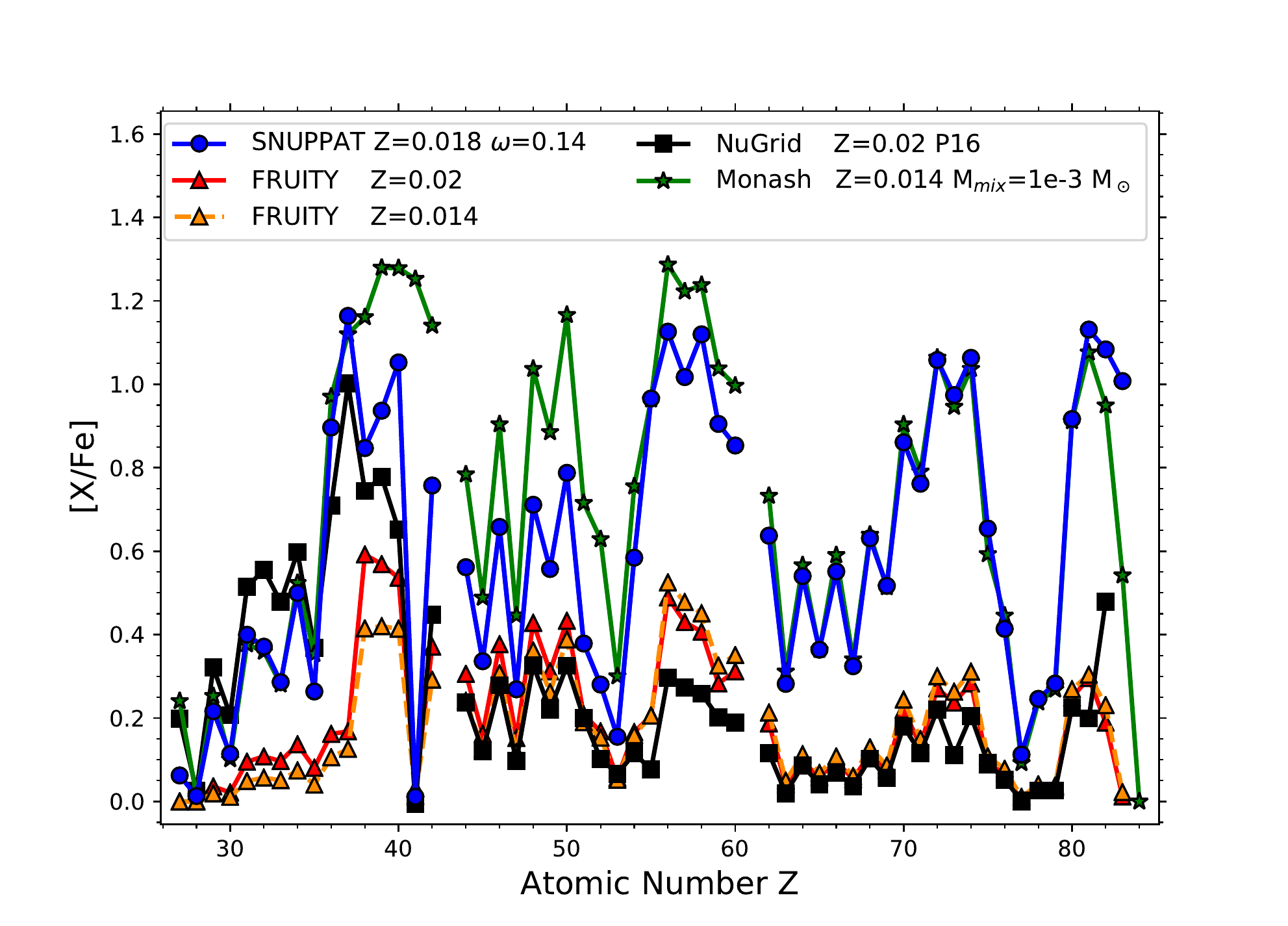}
        \caption{Same as Fig.~\ref{fig:compare3MSun} for 4 M$_\odot$ models, except that we are halving the value of M$_{\text{mix}}$ for the Monash model.}
        \label{fig:compare4MSun}
    \end{figure*}
    
    The nucleosynthesis of 4 M$_\odot$ models is compared in Fig.~\ref{fig:compare4MSun}. There are no models above 3 M$_\odot$ in the work by \citet{Battino2016}, so we compare to the \citet{Pignatari2016} results only. In the 4 M$_\odot$ case, the $^{22}$Ne neutron source is strongly activated in all models except for \textsc{fruity}, where the effects of this neutron source cannot be seen in the surface \textit{s}-process distribution. An explanation for this difference given by \citet{Karakas2016} is that \textsc{fruity} models have fewer TDU episodes, and because the $^{22}$Ne neutron source generally activates more strongly towards the latest TPs, models with fewer TDUs will likely show less $^{22}$Ne neutron source effects in the surface. We find that this difference in number of TDUs exists in our case too, with the \textsc{aton} 4 M$_\odot$ model predicting 14 TDUs with $\lambda > 0.05$ and the 4 M$_\odot$ \textsc{fruity} model at solar metallicity predicting only 8 TDUs above the same threshold.
    
    For this mass, although the \textit{standard} Monash effective $^{13}$C pocket of $10^{-3}$M$_\odot$ is $\sim 4$ times larger than that calculated by \textsc{snuppat} with $\omega = 0.14$ (both pockets having roughly half the mass extent of those found in the 3 M$_\odot$ models), both codes predict abundances of approximately the same order. This is due to the fact that in \textsc{snuppat}, the $^{22}$Ne neutron source activates more strongly than in Monash, providing higher neutron exposure for the nucleosynthesis. The stronger $^{22}$Ne neutron source activation is shown by the increase of the [Rb/Zr] ratio, higher in \textsc{snuppat} than in Monash.
    
    The \textsc{fruity} and NuGrid models both present much lower \textit{s}-process overabundance than Monash and \textsc{snuppat}. In the case of \textsc{fruity}, the $^{22}$Ne neutron source is not activated, which, along with the reduction with increasing initial mass of the $^{13}$C-pocket size, reduces the overall \textit{s} production in \textsc{fruity} between the 3 M$_\odot$ star and the 4 M$_\odot$ star. This is further composed by the fact that \textsc{fruity} has smaller $\lambda$ values than the other codes. In the NuGrid case, we infer from \citet{Pignatari2016} that a 4 M$_\odot$ star should produce an effective $^{13}$C pocket with a mass extent between $5\times10^{-5}$ and $10^{-6}$ M$_\odot$. This is 1 to 2 orders of magnitude lower than what we obtain and is reflected in lower abundances than ours, particularly beyond the first \textit{s}-process peak.
    
    \subsubsection{5 M$_\odot$ models}
    
\begin{figure*}
        \includegraphics[width=\textwidth]{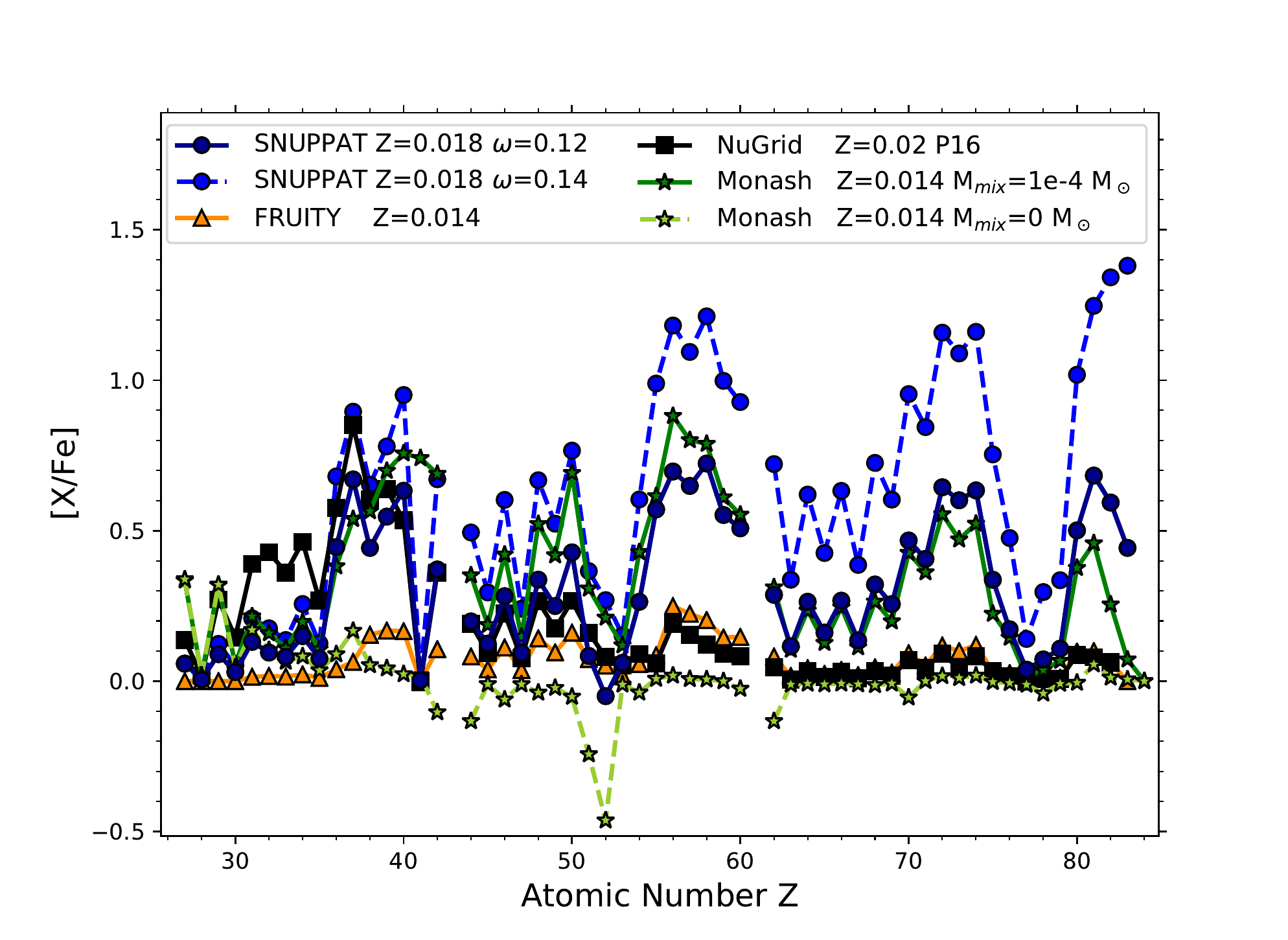}
        \caption{Same as Fig.~\ref{fig:compare3MSun} for 5 M$_\odot$ models, except that we include Monash models with M$_{\text{mix}}$ of $10^{-4}$ M$_\odot$ and $0$. We have also included a model with $\omega = 0.12$ for \textsc{snuppat}, removing one of the \textsc{fruity} models for clarity.}
        \label{fig:compare5MSun}
    \end{figure*}

    For the 5 M$_\odot$ star (Fig.~\ref{fig:compare5MSun}) the main difference relative to the comparisons for the 3 and 4 M$_\odot$ models is that the \textit{standard} Monash simulation has no partial mixing at the bottom of the convective envelope; i.e., no $^{13}$C pocket. This is motivated by the spectroscopic observations of solar metallicity massive AGB stars at the beginning of the TP phase \citep{GarciaHernandez2013}. Such stars display high Li coupled with low Rb and Zr and non-detection of Tc, which suggests the inhibition of the $^{13}$C neutron source. Given that our effective $^{13}$C is not inhibited by the hot TDU mechanism, we have decided to add to the comparisons a 5 M$_\odot$ model from Monash with a $M_\text{mix} = 10^{-4}$ M$_\odot$, which results in an effective $^{13}$C pocket of about $5 \times 10^{-5}$ M$_\odot$, roughly half the $1.10 \times 10^{-4}$ M$_\odot$ effective $^{13}$C pocket of the $\omega = 0.14$ \textsc{snuppat} case. For this reason we have decided to also include the $\omega = 0.12$ calculation, which has an average mass extension of the effective $^{13}$C pocket of about $8.5 \times 10 ^{-5}$ M$_\odot$. This latter \textsc{snuppat} model shows a better agreement with the Monash model. We note that the high Pb abundance in the $\omega = 0.14$ model comes from the strong activation of both neutron sources, which does not happen in any of the other models presented here from \textsc{fruity}, Monash nor NuGrid.
    
    Out of the models presented in Fig.~\ref{fig:compare5MSun}, only those from \textit{non-standard} Monash and \textsc{snuppat} show an production factor in the second \textit{s}-process peak of about $0.5$ dex. The NuGrid model shows a greater first \textit{s}-process peak resulting from a strong $^{22}$Ne neutron source activation. Both the \textsc{fruity} and \textit{standard} Monash models (M$_\text{mix}$ = 0 M$_\odot$) show a small \textit{s}-process overabundance, the consequence of an insignificant size of the effective $^{13}$C pocket.
    
    \subsection{Comparison to the observations}
    \label{sec:CompObserv}
    
    The products of the \textit{s}-process nucleosynthesis in AGB stars can be measured in the AGB stars themselves \citep{Abia2002, GarciaHernandez2006, GarciaHernandez2007, GarciaHernandez2013}, the AGB binary companions enriched by AGB winds (e.g. Ba stars, \citealp{Bidelman1951, deCastro2016}), post-AGB stars (see \citealp{Winckel2003} for a review, \citealp{deSmedt2016}), planetary nebulae (PNe, e.g. \citealp{Schoenberner1981, Sterling2020}), and stardust grains (e.g., \citealp{Zinner2014} for a review). Each source allows us to observe different evolutionary stages of the \textit{s}-process nucleosynthesis, as well as different elements or even isotopes in the grains. The only site not included in this comparison is the post-AGB stars themselves because of the strong low-mass ($< 3$ M$_\odot$), low-metallicity ([Fe/H] $< -0.2$) bias due to the short-lived nature of this evolutionary phase (e.g. \citealp{Winckel2003, Reyniers2007, deSmedt2016}). These lower mass and metallicity fall below those considered here.
    
    We note that the comparisons in this section are limited for two main reasons: i) the first models presented here do not cover the full range of possible progenitor masses (especially the lower masses); and ii) there are several observational limitations; especially the low number of (key) elements that can be measured from stellar/nebular spectra and that strongly varies from C- and O-rich AGBs to PNe and Ba stars.
    
    At solar metallicity, only Rb and upper limits to Zr have been measured in massive O-rich AGB stars, while for C-rich AGBs Rb, Sr, Y, and Zr are mostly available with additional Ba, La, Nd, Sm, and Ce abundances in only 7 solar metallicity C-rich AGBs. From PN nebular spectra Se, Br, Kr, Rb, and Xe are available in two PNe. Finally, for the Ba stars, only 5 \textit{s}-process elements (Y, Zr, La, Ce, and Nd) are mostly available (with La being unreliable; \citealt{Cseh2018} and sometimes Nb and Sm could be measured; e.g., \citealt{Jorissen2019}).
    
    \subsubsection{Carbon stars}
    
\begin{figure}
        \includegraphics[width=\hsize]{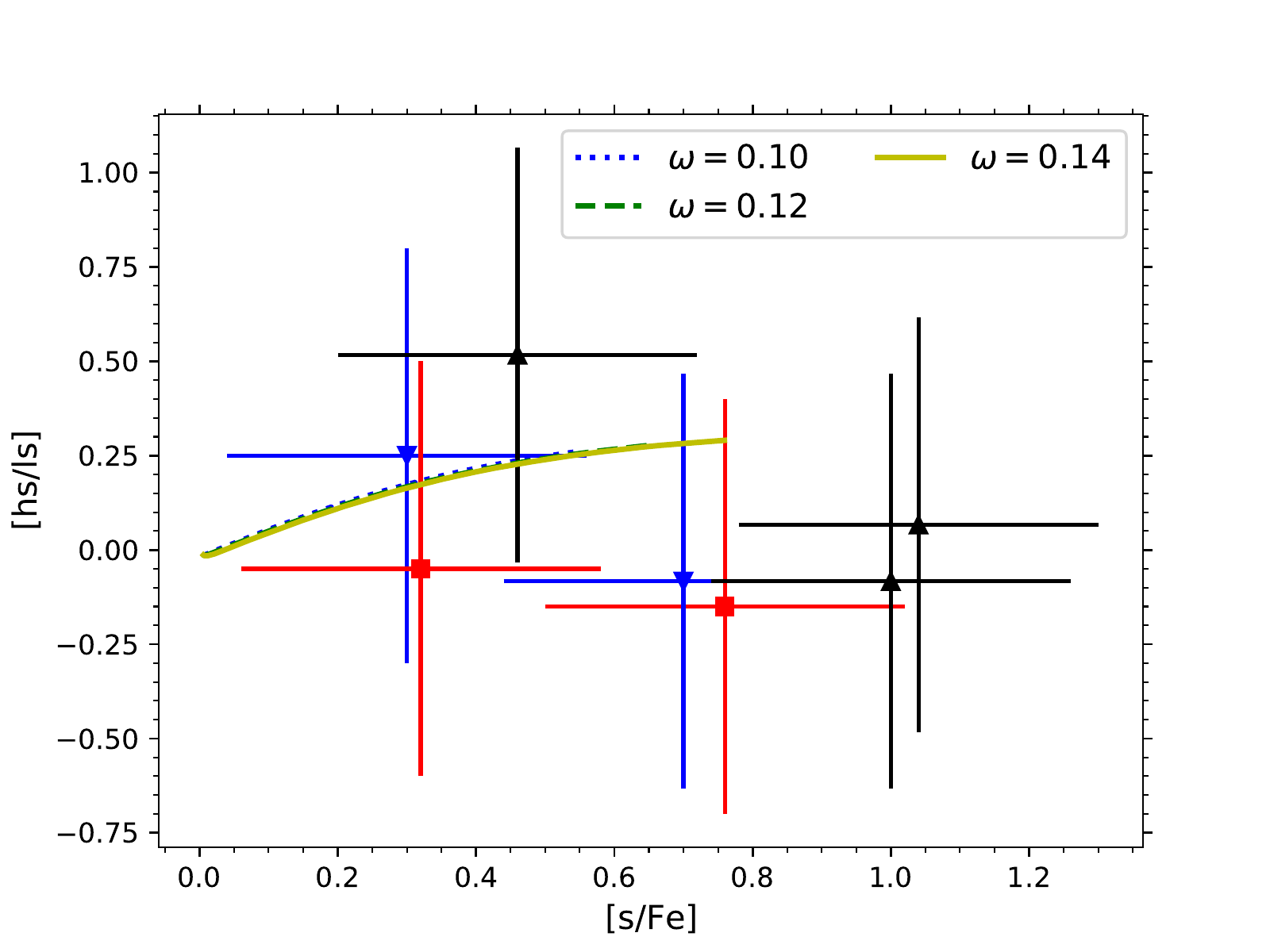}
        \caption{Observed [s/Fe] and [hs/ls] ratios for solar metallicity C stars from \protect\citet{Abia2001, Abia2002} alongside the evolution of ratios for the 3 M$_\odot$ models. The red squares are stars with ``doubtful'' Tc detection. The blue inverted triangles are stars with bolometric magnitude lower than $-5.5$, which should correspond to the more massive among the sample, closer to our 3 M$_\odot$ simulations. The black triangles are stars with bolometric magnitude higher than $-5.5$ and Tc detection. The uncertainties are those given in \protect\citet{Abia2001, Abia2002}.}
        \label{fig:abiaObs}
    \end{figure}

    We consider the carbon stars observed by \citet{Abia2001, Abia2002}. These authors presented observations of a number of \textit{s} elements such as Rb, Sr or Ba as well as Tc in carbon-rich AGB stars. The key measurement of radioactive Tc was made to discriminate between \textit{intrinsic} \textit{s}-process-enhanced stars (where the \textit{s}-process enhancement is produced in the star, henceforth \textit{intrinsic s-star}) and \textit{extrinsic} \textit{s}-process-enhanced stars (\textit{s}-process enhancement obtained after a mass transfer from a companion AGB star, henceforth \textit{extrinsic s-star}), where the presence of Tc indicates the former and its absence the latter. We compare our models to the Ba abundance from this work to calibrate our $\omega$ parameter (we have included the abundances we are comparing to in Table~\ref{tab:abiaVals} of the Appendix). The median [Ba/Fe]\footnote{Although the value is given in [Ba/M], with the metallicity M being an average of Ca, Ti, V, Fe and Co, we have decided to leave the ratio with respect to Fe for consistency. Both values are virtually identical, with a typical [M/Fe] $\sim -0.002$ dex.} abundance of the \textit{intrinsic s-stars} observed by \citet{Abia2002} is $0.75$ with an uncertainty of $0.30$, leaving the three choices for our $\omega$ parameter well within this observational constraint for the 3 M$_\odot$, which is the only among our models that becomes a C star. In Fig.~\ref{fig:abiaObs} we show the [hs/ls] ratio vs the [s/Fe] ratio\footnote{The [s/Fe] ratio is calculated as the average of all the [X/Fe] ratios used for the [hs/Fe] and [ls/Fe] calculations.} as they evolve with time in our 3 M$_\odot$ models along with the data by \citet{Abia2001, Abia2002}. We can see that our results for 5 out of the 7 objects fall within the observational uncertainties, with the other two likely being lower mass stars.
    
     \begin{table}
        \centering
        \caption{The L2 averaged errors between the abundances observed in solar metallicity C-rich AGBs \citep{Abia2001, Abia2002} and the best fitting model in each simulation. The Tc column indicates Tc detection (Y) or doubtful detection (D).}
        \label{tab:abiaObs}
        \begin{tabular}{lccccc}
            \hline
            star & $\omega = 0.10$ & $\omega = 0.12$ & $\omega = 0.14$ & Tc & $M_\text{bol}$\\
            \hline
            AW Cyg & 0.05 & 0.05 & 0.05 & Y & -5.7\\
            S Sct & 0.13 & 0.13 & 0.12 & Y & -4.6\\
            SS Vir & 0.02 & 0.02 & 0.02 & D & ...\\
            SZ Sgr & 0.07 & 0.04 & 0.02 & D & -2.1\\
            U Hya & 0.21 & 0.14 & 0.10 & Y & -4.0\\
            V460 Cyg & 0.03 & 0.01 & 0.01 & Y & -5.8\\
            Z Psc & 0.18 & 0.11 & 0.07 & Y & -4.1\\
            \hline
            Total & 0.69 & 0.50 & 0.39 & & \\
            \hline
        \end{tabular}
    \end{table}
    
    A more comprehensive way to select the best overshooting parameter may be a comparison with the individual elemental abundances (between 7-9 elements, depending on the star) reported by \cite{Abia2001, Abia2002}. Because for AGB stars we are not necessarily witnessing the final composition (see Fig.~\ref{fig:abiaObs}), we compared, for each star, all the model predictions from the 3 M$_\odot$ simulation with the abundances observed. In order to find the best fit, we calculated the average L2 error (i.e., the squared difference) between the surface abundances in the simulation and the observed surface abundances. The values for the best fit for each one of the 3 M$_\odot$ runs are shown in Table~\ref{tab:abiaObs}. Interestingly, this table shows that the best fits are obtained for i) the more luminous stars (M$_\text{bol} \le -5.7$) with Tc (in principle, the higher mass intrinsic AGBs in the sample) or ii) the stars with doubtful Tc, which might be \textit{extrinsic s-stars}. Overall, the overshoot parameter of $\omega = 0.14$ gives the best fit for all the observations.
    
    \subsubsection{Planetary Nebulae}
    
\begin{figure}
        \includegraphics[width=\hsize]{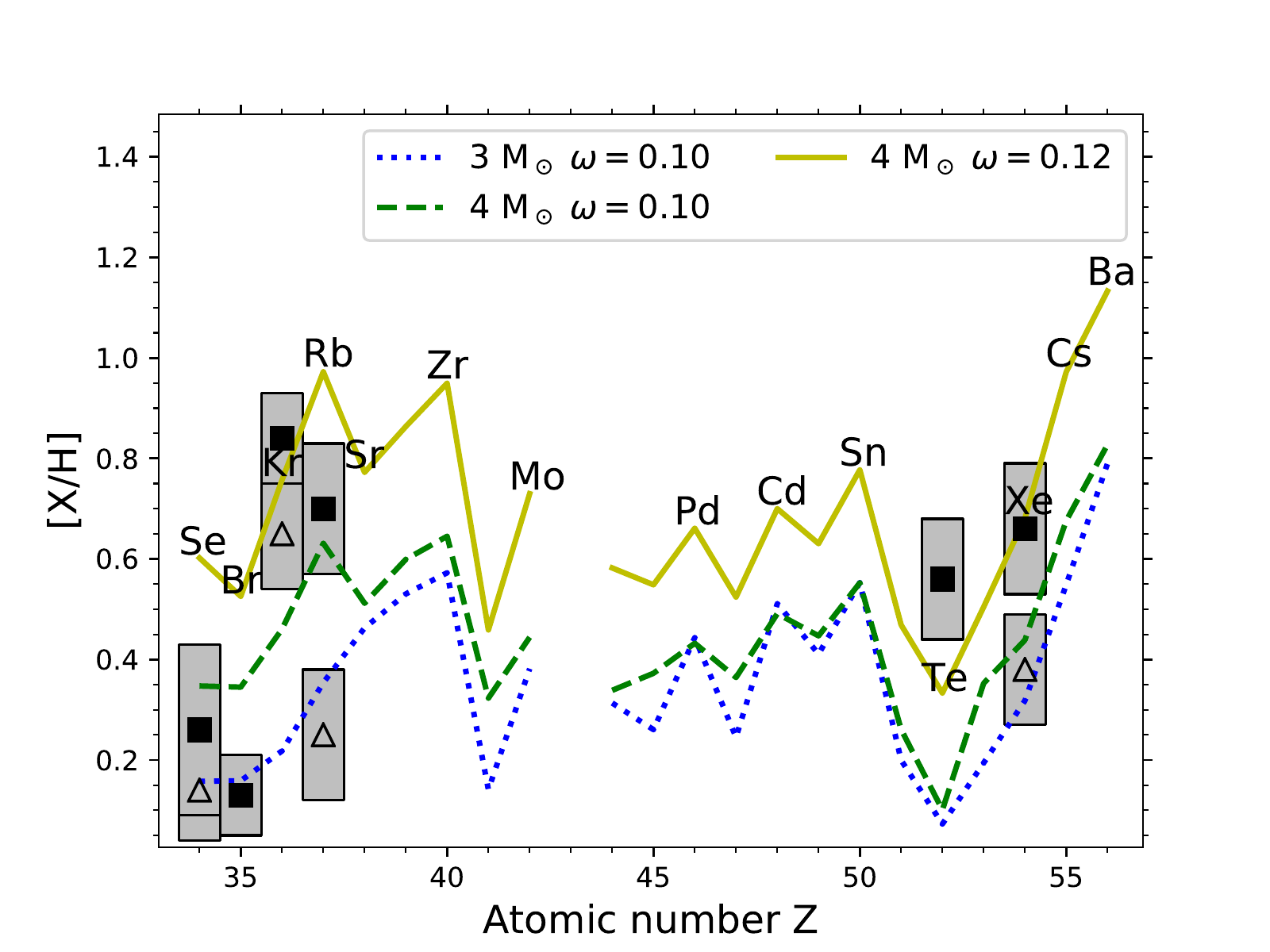}
        \caption{Observed abundances from \protect\citet{MadonnaPhD} of PN NGC 3918 (open triangles) and NGC 7027 (filled squares) alongside the best fitting of our overabundance results (lines and filled dots). In grey are shown the uncertainties for each of the measurements.}
        \label{fig:simonePNe}
    \end{figure}

    In Fig.~\ref{fig:simonePNe}, we show the abundances from \citet{MadonnaPhD} (and references therein, such as \citealp{Madonna2016, Madonna2018}) observed in PNe NGC 3918 and NGC 7027 along with our final selected surface abundance results presented in Fig.~\ref{fig:345MSunResults} for the 3 and 4 M$_\odot$ stars. We have selected these two PNe because they have the most complete and recent solar metallicity PNe observations in regards to neutron-capture elements. As can be seen in the figure, the PNe measurements show an opposite trend for the [Kr/Rb] ratio than our \textit{s}-process models in the range of masses considered here. This ratio may be reversed for less massive stellar models, as is the case with Monash models from 1 to 3 M$_\odot$ (\citealp{MadonnaPhD}). Finally, we point out that one important conclusion of \citet{MadonnaPhD} is that PNe that show evidence of HBB activation have low \textit{s}-process production (see the case of NGC 2440). This observation hints towards either the activation of the hot TDU mechanism that inhibits the $^{13}$C source in more massive stars, in contrast with our models, or the shortening of the AGB lifetime due to a mass transfer scenario.
    
    \subsubsection{Barium stars}
    
    In addition to the previous calibration, we can place our results in context with the work on Ba stars presented by \citet{deCastro2016} and \citet{Cseh2018}. These stars are \textit{extrinsic s-stars}, which means that the observed abundances are diluted with respect to those of the \textit{intrinsic s-stars}. The effects of this dilution are not trivial when transforming from the \textit{intrinsic s-star} abundances to the \textit{extrinsic s-star} abundances. For example, if we consider that a fraction $k$ of the \textit{extrinsic s-star} envelope mass has been added and mixed from the \textit{intrinsic s-star} envelope, the abundances are related by
    \begin{equation}
        \left[X/\text{Fe}\right]_\text{e} = \log_{10}\left(\left(1 - k\right) + k10^{\left[X/\text{Fe}\right]_\text{i}}\right)
        \label{eq:diluteAbund}
    \end{equation}
    where the subindices $i$ and $e$ represent the abundances from the \textit{intrinsic s-star} and \textit{extrinsic s-star}, respectively, and considering that the AGB envelope iron abundance is not significantly modified. This means that, in general, the quantities in Table~\ref{tab:CeY} are affected by the dilution process. However, we can safely assert that, for the values presented in Table~\ref{tab:CeY} (that correspond to a $k = 1$), [Y/Fe] and [Ce/Fe] are more affected than the [Ce/Y] ratio when $k < 1$. This is because the dilution process will affect both Ce and Y while leaving Fe unchanged, due to both AGB and the companion star having the same Fe abundance.
    
    Because the masses of both the observed and the companion stars are estimated to be in the range of 1 to 4 M$_\odot$, we have included our 3 and 4 M$_\odot$ models in the comparison.
    
    We can compare our results of Table~\ref{tab:CeY} to Fig.~6 of \citet{Cseh2018}, where the bulk of [Ce/Y] values at solar metallicity range from $-0.4$ dex to below $0.2$ dex. At this metallicity, our [Ce/Y] ratios are above the trend shown by the observations, slightly above the 4 M$_\odot$ \textsc{fruity} and Monash ratios. If we consider dilution in the range of $0.1 < k < 0.5$, our results move towards the bulk of the measurements, although remaining with the positive [Ce/Y] abundances.
    
    \begin{table}
        \centering
        \caption{Average of [Y/Fe] and [Ce/Fe] as well as ratio [Ce/Y] by mass and overshoot parameter $\omega$ from our simulations.}
        \label{tab:CeY}
        \begin{tabular}{lccr}
            \hline
            Mass (M$_\odot$) & $\omega$ & $\frac{1}{2}$([Y/Fe] + [Ce/Fe]) & [Ce/Y]\\
            \hline
            3 & 0.10 & 0.51 & 0.23 \\
              & 0.12 & 0.61 & 0.25 \\
              & 0.14 & 0.72 & 0.27 \\
            4 & 0.10 & 0.42 & 0.27 \\
              & 0.12 & 0.58 & 0.33 \\
              & 0.14 & 1.03 & 0.18 \\
            \hline
        \end{tabular}
    \end{table}
    
     \begin{table}
        \centering
        \caption{Abundances observed in the 3 M$_\odot$ Ba star BD-142678 \citep{deCastro2016} in comparison with the predicted ones from our best fit 3 and 4 M$_\odot$ models with $\omega = 0.14$. The $k$ values (those used in Eq.~\ref{eq:diluteAbund}) and the average L2 errors are also listed.}
        \label{tab:BaStars}
        \begin{tabular}{lccc}
            \hline
            Value & BD-142678 & 3 M$_\odot$ & 4 M$_\odot$ \\
            \hline
            {[Y/Fe]} & 1.02 & 0.64 & 0.98 \\
            {[Zr/Fe]} & 0.85 & 0.72 & 1.07 \\
            {[La/Fe]} & 0.91 & 0.94 & 1.09 \\
            {[Ce/Fe]} & 0.87 & 0.94 & 1.20 \\
            {[Nd/Fe]} & 0.87 & 0.78 & 0.96 \\
            k & - & 1.00 & 0.60 \\
            L2 error & - & 0.03 & 0.02 \\
            \hline
        \end{tabular}
    \end{table}
    
    Similarly to the C-rich AGB stars, we have compared the individual abundances observed in the solar metallicity Ba star BD-142678 \citep{deCastro2016} with those predicted by our best fit 3 and 4 M$_\odot$ models with $\omega = 0.14$. We have selected this particular Ba star because its progenitor mass is inferred to be ~3 M$_\odot$ (i.e., closer to our models), when compared with both Monash and \textsc{fruity} model predictions diluted to fit the observed [Ce/Fe] ratio \citep{Cseh2020}. The comparison is shown in Table~\ref{tab:BaStars}, where we also list the dilution factor $k$ needed to get the best match as well as the average L2 errors. Note that for the 3 M$_\odot$ star model we report a value of $k = 1$, which corresponds to the maximum possible value for $k$, not necessarily the value with the best fit, which needs $k > 1$. Such a value would require yields that are not attainable by diluting our model. If we try to fit the observed [Ce/Fe] ratio, then our best fit 3 M$_\odot$ model would need a dilution of $k = 0.83$, which is very close to the dilution value of $0.93$ for the 3 M$_\odot$ \textsc{fruity} model fitted for this Ba star by \cite{Cseh2020}.
    
    \subsubsection{Massive AGBs}
    
    Another test of our simulations is a comparison with the observed [Rb/Fe] abundances in massive solar metallicity AGB stars. Some of the first observations of the [Rb/Fe] ratio (with a typical large uncertainty up to $\pm0.8$ dex) were obtained by \citet{GarciaHernandez2006}, who reported values ranging from $-1.0$ to $2.5$ dex. More recent work by \citet{Zamora2014} and \citet{PerezMesa2017} narrows this spread to values between $-0.7$ and $1.3$ dex with an uncertainty of $\pm0.7$ dex. As can be seen from Fig.~\ref{fig:345MSunResults}, our [Rb/Fe] values fall well within those constrains for our more massive models. However, the [Rb/Zr] ratios reported in \citet{Zamora2014} and \citet{PerezMesa2017} range from $-0.3$ dex to $0.7$ dex with uncertainties between $0.7$ and 1 dex, while in our 4 and 5 M$_\odot$ models we find that the [Rb/Zr] ratio remains between $-0.1$ and $0.1$ dex. This is consistent with the ratios found by other groups. This gap between models and observation has been discussed in the literature before, and may be attributed to the different dust condensation temperatures of Zr and Rb, with Zr being removed from the gas at higher temperatures than Rb (see e.g. \citealp{GarciaHernandez2009, vanRaai2012}).
    
    \subsubsection{Dust grains}
    
    Finally, we compare to isotopic ratios of \textit{s}-process elements measured in meteoritic silicon carbide (SiC) \citep{Lugaro2018Grains}. Most of these grains formed in AGB C-rich stars such as our 3 M$_\odot$ model (see \citealp{DellAgli2015} and references therein) and carry information on the \textit{s}-process nucleosynthesis of the stars where they formed (e.g., see \citealp{Liu2014, Lugaro2014Grains}). Certain isotopes, such as those of Zr, strongly depend on the \textit{s}-process neutron density due to a branching point in $^{95}$Zr, as well as both the neutron exposure and \textit{s}-process production given by the effective $^{13}$C pocket mass fraction and mass extension. By comparing our results (Table~\ref{tab:grains}) with Fig.~2 of \citet{Lugaro2018Grains}, we find that our 3 M$_\odot$ models cover some of the observed values of $\delta^{91}$Zr/$^{94}$Zr\footnote{The $\delta$ notation is the permil variation compared to the solar ratios, which are $\delta = 0$ by definition.}, $\delta^{92}$Zr/$^{94}$Zr, and $\delta^{96}$Zr/$^{94}$Zr, with larger $\omega$ values agreeing better with the observed ratios. In fact, our 3 M$_\odot$ models are very similar to the ``M2.z2m2'' points shown in Fig.~14 of \citet{Battino2016}, just shifted left towards the bulk of the measurements. This is expected because our high [hs/ls] for this 3 M$_\odot$ model is produced by a high neutron exposure due to the large effective $^{13}$C mass fraction, which is also tied to a high neutron density for the $^{13}$C neutron source and the partial activation of the $^{22}$Ne neutron source due to the higher temperature.
    
    \begin{table}
        \centering
        \caption{Final permil variation with respect to the solar ratio ($\delta$ notation) for surface isotopic ratios of Zr in the 3 M$_\odot$ model calculated in this work.}
        \label{tab:grains}
        \begin{tabular}{lcccr}
            \hline
            $\omega$ & $\delta^{90}$Zr/$^{94}$Zr & $\delta^{91}$Zr/$^{94}$Zr & $\delta^{92}$Zr/$^{94}$Zr & $\delta^{96}$Zr/$^{94}$Zr\\
            \hline
            0.10 & -291 & -193 & -129 & -501 \\
            0.12 & -318 & -209 & -137 & -536 \\
            0.14 & -340 & -218 & -143 & -556 \\
            \hline
        \end{tabular}
    \end{table}

\section{Conclusions}
\label{sec:Conclusions}
    
    We have presented new \textit{s}-process nucleosynthesis predictions based on the \textsc{aton} stellar evolutionary models at solar metallicity (Z $= 0.018$) for stars of 3, 4, and 5 M$_\odot$. These results show that
    
    1 - In the more massive AGB stars ($\gtrsim 4$ M$_\odot$), our overshoot prescription does not find the effective $^{13}$C pocket inhibition described by \citet{Goriely2004}. This is because the overshoot scheme adopted in \textsc{snuppat} is advective instead of diffusive. Therefore, for simulations where hot TDU episodes are expected to inhibit the effective $^{13}$C pocket, this overshoot prescription should not be used. If a diffusive overshoot scheme is implemented instead, the $^{13}$C pocket may be inhibited in our 4 and 5 M$_\odot$ cases because in \textsc{aton} both those stars experience hot TDU episodes for most of their lifetimes. This may be also the reason for the slight disagreement with the PNe observations. In particular, the lower \textit{s}-process yield of some of them. This needs to be investigated in future work.
    
    2 - The envelope overshoot parameter of $\omega = 0.14$ reproduces well the observed abundance of [Ba/Fe] for solar metallicity carbon stars. This parameter also yields the best agreement with other standard \textit{s}-process nucleosynthesis models, specially for the first and second \textit{s}-process peaks. We note that this overshoot parameter is a consequence of post-processing models that have inconsistent overshoot with the structure models. If the overshoot feedback was included in the structure models, it is likely that we would obtain a smaller overshoot parameter.
    
    3 - Due to the high $^{12}$C intershell mass fraction our models produce a higher neutron exposure and [hs/ls] value than the Monash and \textsc{fruity} models of the same mass. For stars of $< 4$ M$_\odot$ we find very similar results to NuGrid in spite of our different approaches to the overshoot mechanism.
    
    4 - The \textsc{aton} models achieve a high intershell maximum temperature, efficiently activating the $^{22}$Ne neutron source for the 4 and 5 M$_\odot$ models and producing high [Rb/Fe] ratios in accordance with observations of massive AGB stars. Although the [Rb/Zr] ratios are much lower than those observed, they remain in agreement with the other \textit{s}-process nucleosynthesis models.
    
    5 - Some of the isotopic Zr dust grain measurements can be reproduced by our C-rich 3 M$_\odot$ model, although there is still some disagreement with the $\delta^{96}$Zr/$^{94}$Zr value for the bulk of the measurements, likely attributed to the high neutron exposure occurring in our simulations. The $\omega$ necessary to reproduce these results could be lower if the feedback between overshoot and structure was taken into account.
    
    We have also introduced a new explicit integration method, the Patankar-Euler-Deuflhard solver, and we have shown that an implementation in NuPPN is as accurate as the Bader-Deufhlard solver and faster than both the Bader-Deufhlard and Backwards Euler implementations for the same tolerance in a He-burning and $^{13}$C pocket trajectories.
    
    The natural extension of this work is the exploration of \textit{s}-process abundances produced by \textsc{aton} models for subsolar and supersolar metallicities, as well as extend our mass range to lower and higher initial stellar masses. Furthermore, we will experiment with different overshoot and convective mixing prescriptions to understand the effects that both hot TDU episodes and the HBB can have on the \textit{s}-process overabundances depending on the numerical modelling. Finally, we will also explore more systematically the effects of the feedback between overshoot and stellar structure on \textit{s}-process nucleosynthesis.

\begin{acknowledgements}
    This work has been supported by the European Research Council (ERC-2016-CO Grant 724560) and by the US Department of Energy LDRD program through the Los Alamos National Laboratory. Los Alamos National Laboratory is operated by Triad National Security, LLC, for the National Nuclear Security Administration of U.S. Department of Energy (Contract No. 89233218NCA000001). DAGH acknowledges support from the State Research Agency (AEI) of the Spanish Ministry of Science, Innovation and Universities (MCIU) and the European Regional Development Fund (FEDER) under grant AYA-2017-88254-P. CD acknowledges support from the Lend\"ulet-2014 Programme of the Hungarian Academy of Sciences. We thank the anonymous referee for their useful comments that improved the quality of this work.
\end{acknowledgements}
\bibliographystyle{aa} 
\bibliography{aylrefs} 

\appendix

\section{Test of overshoot schemes for the inhibition of the effective $^{13}$C pocket}
\label{sec:appInhibition}  

\begin{figure}[hbt]
    \begin{center}
        \includegraphics[width=0.4\textwidth]{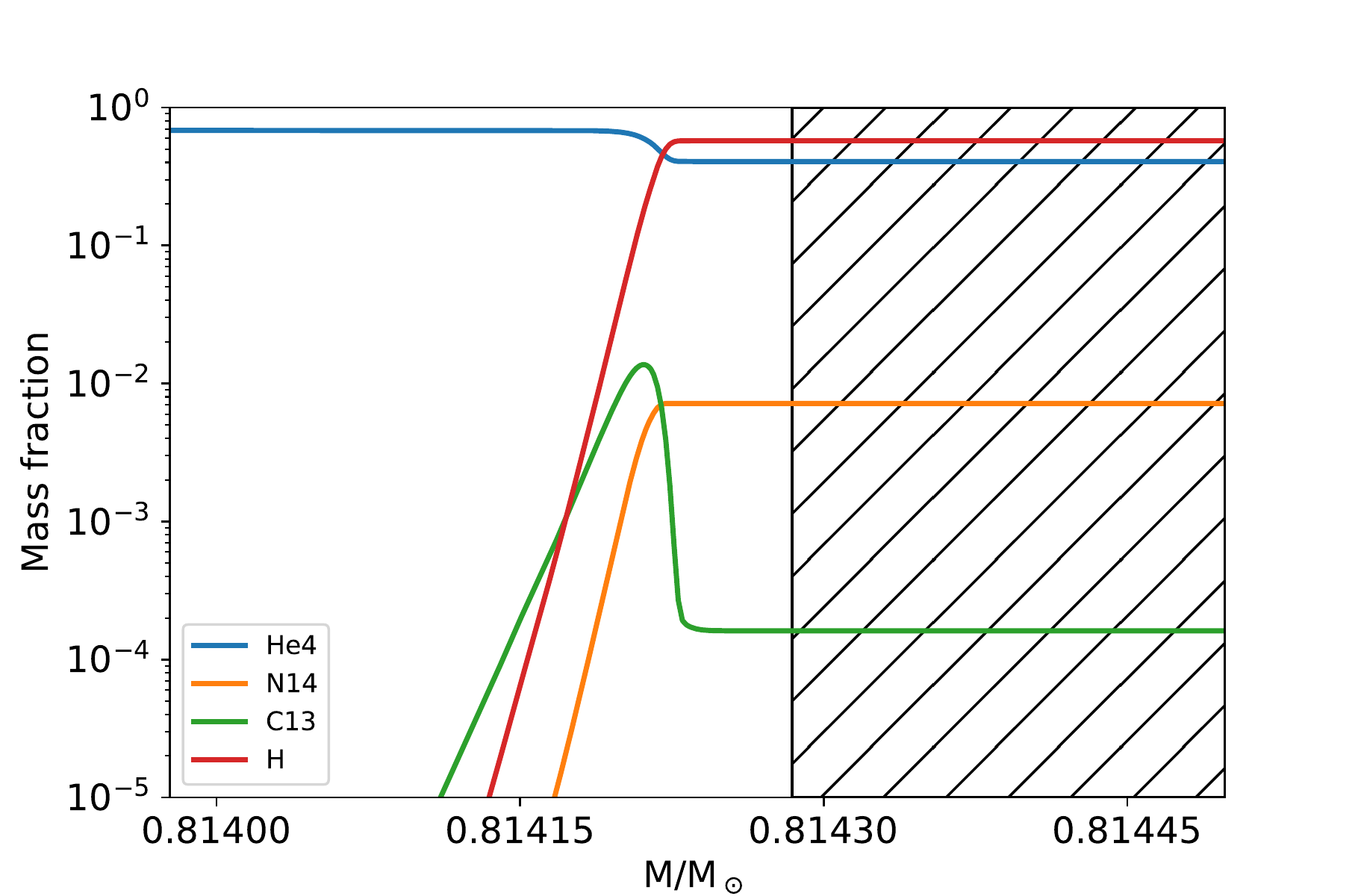}
        \includegraphics[width=0.4\textwidth]{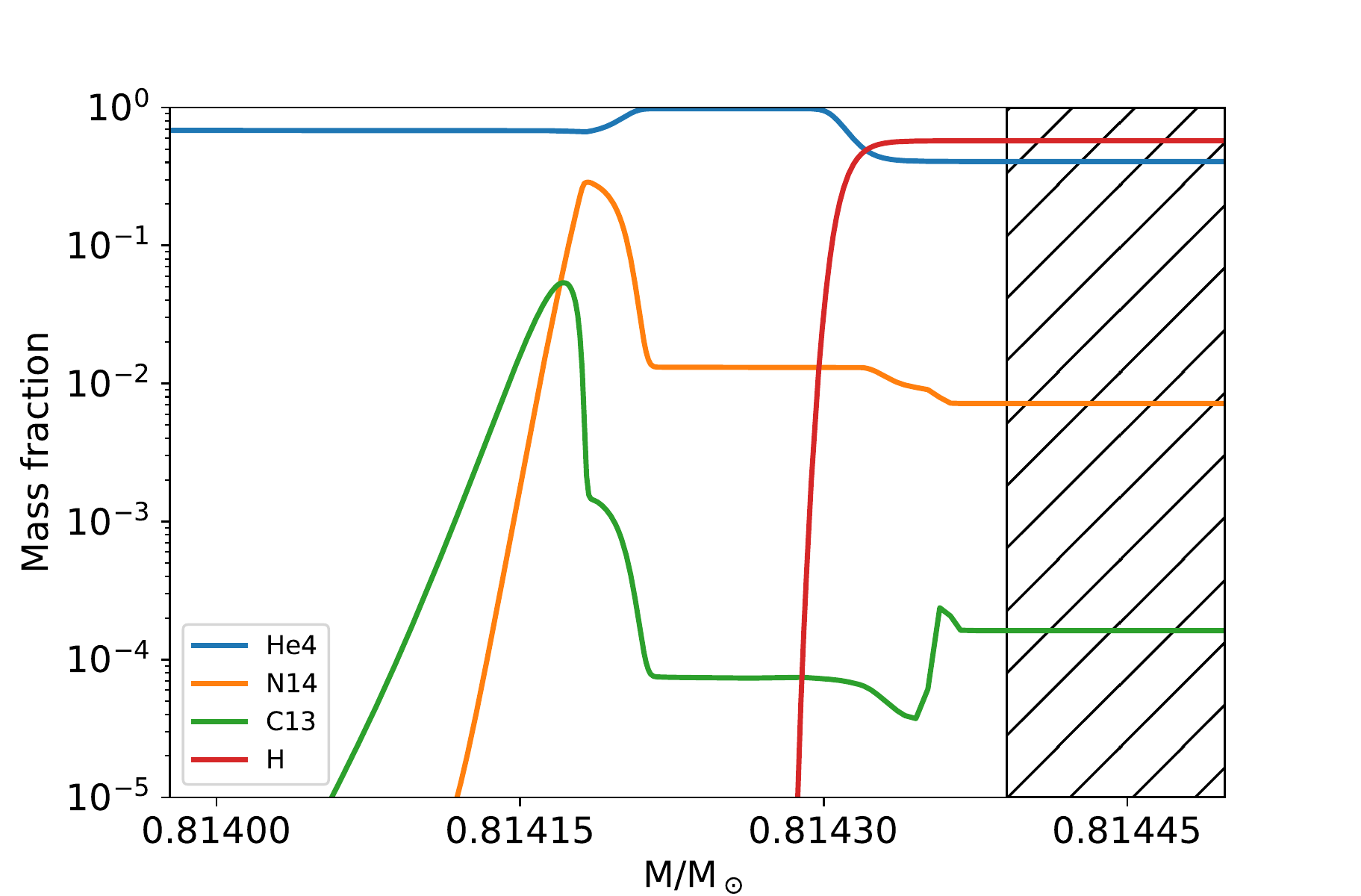}
    \end{center}
    \caption{Chemical profiles of one of the hot TDU in our 4 M$_\odot$ model with advective mixing. At the top, prior to the deepest extent of the TDU, and at the bottom afterwards. The non-inhibited effective C13 source is clear in this case.}
    \label{fig:4Msun_adv}
\end{figure}

\begin{figure}[hbt]
    \begin{center}
        \includegraphics[width=0.4\textwidth]{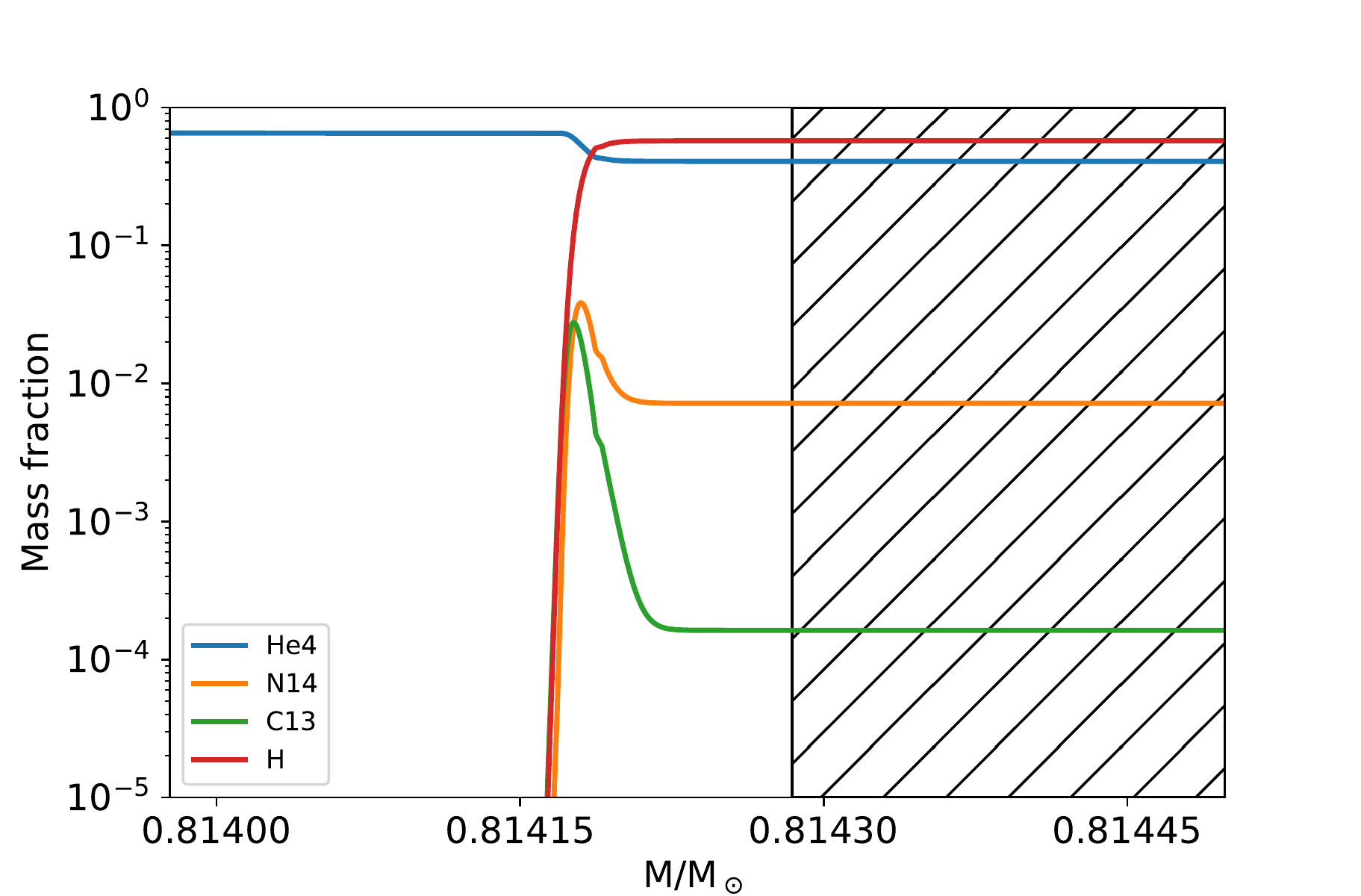}
        \includegraphics[width=0.4\textwidth]{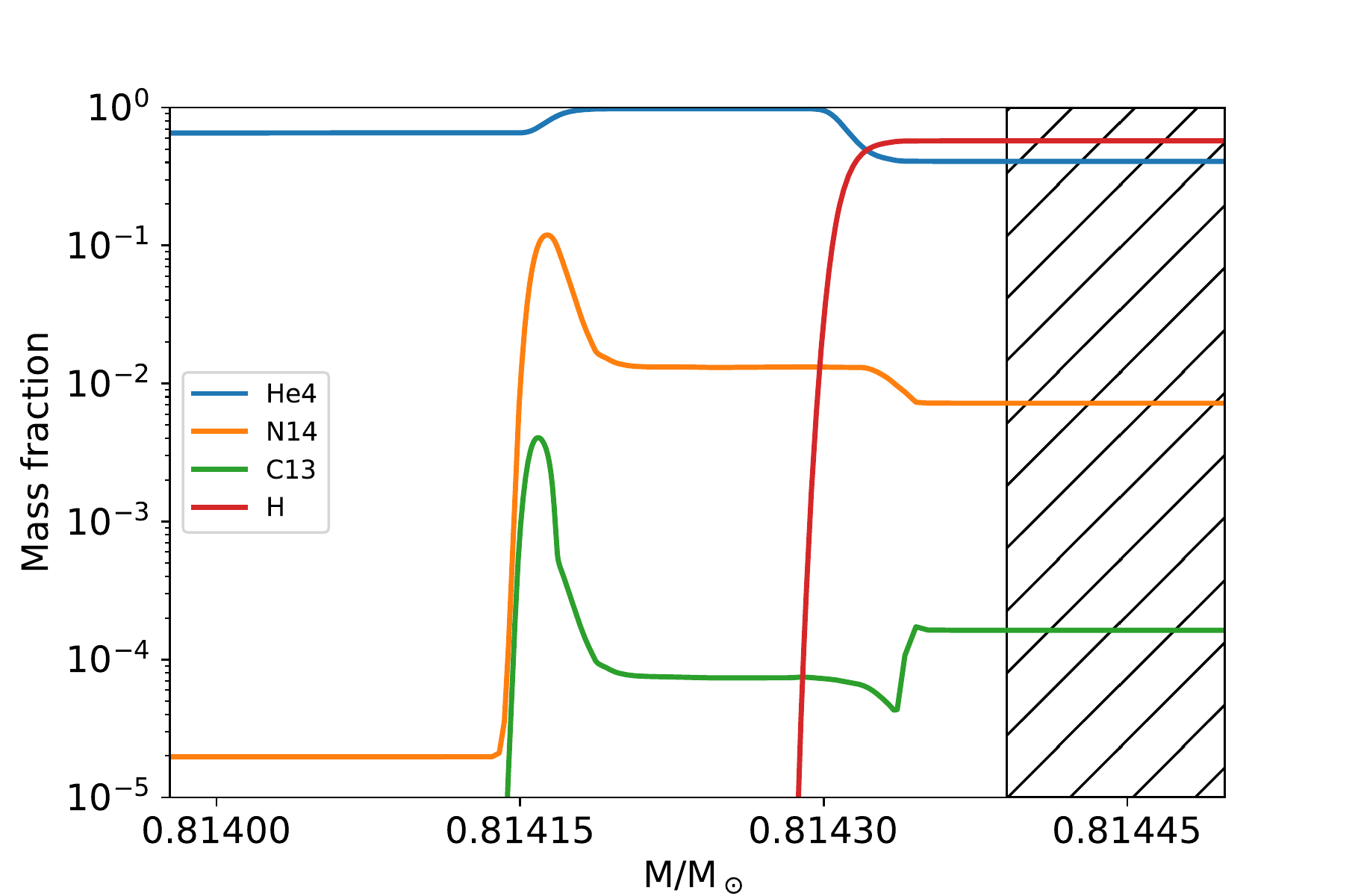}
    \end{center}
    \caption{Same as Fig.~\ref{fig:4Msun_adv} with a diffusive mixing. In this case the coupling of mixing and burning through operator-splitting inhibits the C13 neutron source by placing the C13-pocket inside of the N14-pocket.}
    \label{fig:4Msun_diff}
\end{figure}

\section{Relevant data}

\begin{table}[hbt]
    \centering
    \caption{Observed values reported by \citet{Abia2001, Abia2002} for the 7 solar metallicity stars that we compare with in this work.}
    \label{tab:abiaVals}
    \begin{tabular}{lccccccccc}
        \hline
        Name & [Rb/Fe] & [Sr/Fe] & [Y/Fe] & [Zr/Fe] & [Ba/Fe] & [La/Fe] & [Ce/Fe] & [Nd/Fe] & [Sm/Fe] \\
        \hline
        AW Cyg & 0.2 & 0.4 & 0.3 & 0.5 & 0.0 & 0.3 & - & 0.5 & - \\
        S Sct & 0.5 & 0.8 & 0.7 & 0.5 & 0.2 & 0.1 & - & 0.2 & - \\
        SS Vir & - & 0.0 & 0.5 & 0.4 & 0.3 & 0.4 & - & 0.3 & - \\
        SZ Sgr & 0.1 & 0.4 & 0.9 & 0.8 & 0.8 & 0.9 & - & 0.9 & 0.6 \\
        U Hya & 0.5 & 0.8 & 1.3 & 1.1 & 1.1 & 0.9 & 0.6 & 1.0 & 0.7 \\
        V460 Cyg & 0.4 & 0.5 & 0.7 & 0.8 & 0.8 & 0.7 & - & 0.8 & 0.4 \\
        Z Psc & 0.6 & 0.9 & 1.0 & 1.0 & 1.0 & 1.1 & 0.6 & 0.9 & 0.8 \\
        \hline
    \end{tabular}
\end{table}

\end{document}